\documentclass[aps,prd,12pt,nofootinbib]{revtex4}

\usepackage{graphics}

\usepackage{epsfig}
\usepackage{amsmath}	
\usepackage{amssymb}    

\newcommand{\be}{\begin{eqnarray}}
 \newcommand{\ee}{\end{eqnarray}}

 \newcommand{\bra}[1]{\left\langle #1 \right|}
 \newcommand{\ket}[1]{\left| #1 \right\rangle}

\begin{document}


\pagestyle{empty}

\title{Wilson lines - color charge densities correlators and the production of $\eta'$ in the CGC for $pp$ and $pA$ collisions}


\author{Fran\c{c}ois Fillion-Gourdeau}
\email{ffillion@hep.physics.mcgill.ca}

\author{Sangyong Jeon}
\email{jeon@physics.mcgill.ca}

\affiliation{Department of Physics, McGill University, 3600 University Street, Montreal, Canada H3A 2T8}

\date{\today}

\begin{abstract}
We compute the inclusive differential cross section production of the pseudo-scalar meson $\eta'$ in high-energy proton-proton ($pp$) and proton-nucleus ($pA$) collisions. We use an effective coupling between gluons and $\eta'$ meson to derive a reduction formula that relates the $\eta'$ production to a field-strength tensor correlator. For $pA$ collisions we take into account saturation effects on the nucleus side by using the Color Glass Condensate formalism to evaluate this correlator. We derive new results for Wilson line - color charges correlators in the McLerran-Venugopalan model needed in the computation of $\eta'$ production.  The unintegrated parton distribution functions are used to characterize the gluon distribution inside protons. We show that in $pp$ collisions, the cross section depends on the parametrization of unintegrated parton distribution functions and thus, it can be used to put constraints on these distributions. We also demonstrate that in $pA$ collisions, the cross section is sensitive to saturation effects so it can be utilized to estimate the value of the saturation scale. 
\end{abstract}

\maketitle


\pagestyle{plain}


\section{Introduction}

One of the main challenges in high energy hadronic collisions is the understanding of particle production. The theoretical description of these complex phenomena involves both many-body physics and the theory of strong interactions. There are many approaches that aim toward a better comprehension of these topics. One of the most successful technique is perturbative quantum chromodynamics (pQCD) where one studies the limit where the coupling constant is small and where the usual loop expansion can be used in principle. Because QCD is an asymptotically free theory, this happens when the exchanged momenta are large compared to the QCD scale $\Lambda_{\rm{QCD}}$. Even in that regime however, physical observables computed using this machinery suffer from infrared divergences that spoil the naive perturbative expansion. These have to be resummed and this leads to various factorization formalisms like the collinear factorization and the $k_{\perp}$-factorization. In collinear factorization, meaningful physical quantities are obtained in terms of parton distribution functions (PDF). These distributions characterize the non-perturbative (large distance) physics and have to be determined experimentally from a fit to structure functions. This formalism can be applied on system where the typical exchanged momentum $Q$ is hard, which means that it satisfies the inequality given by $\Lambda_{\rm{QCD}}^{2} \ll Q^{2} \sim s$ where $\Lambda_{\rm{QCD}} \approx 0.2 \; \mbox{GeV}$ is the QCD scale and $\sqrt{s}$ is the center of mass energy. Note that $Q^{2}\sim M_{\perp}^{2} \equiv M^{2}+k_{\perp}^{2}$, which means that unless the mass of the produced particle $M$ is of the order of $\sqrt{s}$, the collinear factorization is valid only for very large $k_{\perp}$. This can be relaxed in the $k_{\perp}$-factorization formalism which considers semihard collisions, meaning that the typical exchanged momentum obeys $\Lambda_{\rm{QCD}}^{2} \ll Q^{2} \ll s$. The resummation implemented in this approach takes care of large contributions that look like $\left[ \ln(Q^{2}/\Lambda_{QCD}^{2}) \alpha_{s} \right]^{n}$, $\left[ \ln(Q^{2}/\Lambda_{QCD}^{2}) \ln(1/x) \alpha_{s} \right]^{n}$ and $\left[  \ln(1/x) \alpha_{s} \right]^{n}$ \cite{Collins:1991ty,Catani:1990eg,Gribov:1984tu,Kuraev:1977fs}. This technique is used successfully to compute the production of many kinds of particles in high energy proton-proton collisions like heavy quarks \cite{Collins:1991ty,Luszczak:2005cq,Lipatov:2001ny,Zotov:2003cb,Jung:2001rp} and in a number of other processes (see \cite{Andersson:2002cf,Andersen:2003xj,Andersen:2006pg} for reviews of many applications). It is also used for predictions of Higgs boson production \cite{Luszczak:2005xs,Lipatov:2005at}. In this article, we use this formalism to compute the $\eta'$ inclusive cross section at the RHIC energy. 

When there is a nucleus involved in a collision at very high energy, there are new effects not included in the previous approaches due to the high density of gluons resulting from the emission enhancement at small-$x$ (where $x$ is the momentum fraction). These effects introduce a new scale $Q_{s}$ called the saturation scale at which the probability of having interactions between gluons of the same nucleus becomes important. At this transverse momentum scale, the gluons recombine and this slows down the growth of partons at smaller $x$. A naive estimation of $Q_{s}$ shows that it depends on $x$ and the number of nucleons $A$ like $Q_{s}^{2} \sim A^{\delta} x^{-\lambda}$ \cite{Iancu:2002xk,Iancu:2003xm} so at small enough $x$ or large enough $A$, the saturation scale is hard ($Q_{s}^{2} \gg \Lambda_{\rm{QCD}}^{2}$). When the typical exchanged momentum is smaller than the saturation scale such as $\Lambda_{\rm{QCD}}^{2} \ll Q^{2} \leq Q_{s}^{2} \ll s$, saturation effects have to be taken into account even if the system is still in the perturbative regime. This can be achieved in the Color Glass Condensate (CGC) formalism which is a semi-classical effective theory where the non-linearities are dealt with by solving exactly the Yang-Mills equation of motion. This takes care of gluon recombinations and introduces the effects of saturation in observables.

In this article, we are using the CGC to compute the inclusive differential cross section of $\eta'$ meson in $pA$ collisions at the RHIC energy ($\sqrt{s}=200 \; \mbox{GeV}$). The main goal of this work is to look at the effect of saturation in $\eta'$ production to validate the CGC approach and estimate the value of the saturation scale by comparing our predictions with experimental data. The $\eta'$ is a pseudoscalar meson with a mass of $M=0.957 \; \mbox{GeV}$, a decay width of $0.203 \; \mbox{MeV}$ and quantum numbers of $I^{G}(J^{PC})=0^{+}(0^{-+})$ \cite{Yao:2006px}. One of the most important features of $\eta'$ is that it couples to the QCD anomaly \cite{Hooft:1976up,Witten:1979vv}. One way to implement and model this physics is by introducing an effective interaction between gluons and $\eta'$ mesons. This was done in \cite{Atwood:1997bn}, where the authors are proposing a vertex that couples two gluons and a $\eta'$ meson ($gg\eta'$) to explain B mesons decay ($B \rightarrow \eta'+ X$). This vertex contains a form factor that depends generally on gluons and $\eta'$ momenta and that can be related to the $\eta'$ wave-function. The structure of this vertex was investigated thoroughly using various techniques like the hard scattering and the running coupling approaches. \cite{Ali:2003kg,Ali:2000ci,Muta:1999tc,Ahmady:1998mi,Agaev:2002ek,Kroll:2002nt}. We use these results on the gluons-$\eta'$ coupling to study the $\eta'$ production mechanism based on gluon fusion.  

The $\eta'$ production in $pp$ collisions at high energy was studied in \cite{Szczurek:2006bn} where the exclusive cross section for the diffractive process $p+p \rightarrow p+p+\eta'$ is computed. In our study, we focus on the inclusive production mechanism $p+p \rightarrow \eta' +X$ which shares similar features with this previous analysis. The first attempt to compute $\eta'$ production in high energy $pA$ collisions was done by one of the present author in \cite{JalilianMarian:2001bu}. In this study, the collinear factorization is used to compute the cross section at RHIC by including intrinsic transverse momentum in the PDF with a Gaussian distribution. Based on physical arguments, the width of the Gaussian, which represents the typical transverse momentum of gluons inside the nucleus, is chosen to be $Q_{s}$. The authors show that the $\eta'$ production is sensitive to the saturation scale implemented in this way. However, they acknowledge that their calculation can be improved because they use the collinear formalism outside its range of validity. The goal of this article is to revisit the $\eta'$ production with a more rigorous approach by doing a full CGC computation that includes recombination effects more realistically. 

The computation of meson production in the CGC was undertaken in the past using mostly an ``hybrid'' approach where the proton and the nucleus are described by the collinear factorization and the CGC respectively. In this formalism based on pQCD-like techniques, the fragmentation function of collinear factorization is convoluted with the gluon or quark cross section computed in the CGC formalism. This is suitable for well-known mesons like pions for which a wealth of experimental data have been measured and for which fragmentation functions are well-known. Pion production for $pA$ collisions is computed in \cite{Dumitru:2005gt,Dumitru:2005kb,Tuchin:2007pf} using this methodology. Contrary to pions, the data in high energy hadronic collisions for $\eta'$ is scarce, so another approach is required to take care of hadronization effects and internal structure of the $\eta'$ meson. In \cite{FillionGourdeau:2007ee}, an effective theory is used to estimate the tensor meson production in $pp$ collisions. We use a similar approach in this article where the interaction between gluons and $\eta'$ is described by an effective theory. As discussed previously, we include these effects in an effective vertex that includes a form factor. As will be shown in this article, this can be implemented easily in the CGC formalism. 

We consider only the case of $pp$ and $pA$ collisions at RHIC. For nucleus-nucleus ($AA$) collisions, the total number of $\eta'$-mesons produced by semihard collisions in the first instants (for $t<1 \; \mbox{fm/c}$) should be important and the saturation effects would also be present. Experimentally however, they cannot be detected because most of them decay inside the medium created by the collision. This is because the $\eta'$ mean lifetime, which is about $t_{\eta'} \approx 4.93 \; \mbox{fm/c}$, is smaller than the time where the medium exists, which is from $1 \; \mbox{fm/c}$ up to $10 \; \mbox{fm/c}$. Moreover, by considering $pp$ and $pA$, we avoid all the complications that would result from the creation of the medium which include the understanding and modelling of the quark-gluon plasma properties. Finally, there are analytical solutions for the gauge field in $pp$ and $pA$ collisions, while the analytical solution in $AA$ is still elusive. For these reasons, our present analysis is only applied to $pp$ and $pA$ collisions.

This article is organized as follows. In section \ref{sec:eff_th} we describe the effective vertex used throughout the rest of the article. In section \ref{sec:prod_CGC}, we show how to compute $\eta'$ production in the CGC for $pA$ and $pp$ collisions. We start by deriving a reduction formula that relates the cross section to a correlator of field-strength tensors. This correlator is then evaluated to leading order in $pA$ collisions and the result can be interpreted in terms of physical processes. We also show how the $k_{\perp}$-factorized cross section for $pp$ collisions can be recovered in the low density limit of $pA$ cross section. In section \ref{sec:averages}, we compute the correlators appearing in the expression of the $pA$ cross section using the McLerran-Venugopalan model. In section \ref{sec:num_eval}, we evaluate numerically the cross section for $pA$ and $pp$ and discuss the range of validity of our computation. Sections III and IV contain a lot of technical details. The reader interested in results can jump directly to section V.

Throughout the article, we use both light-cone coordinates defined by
\begin{eqnarray}
p^{+} &=& \frac{p^{0} + p^{3}}{\sqrt{2}} \; ; \; 
p^{-} = \frac{p^{0} - p^{3}}{\sqrt{2}}
\end{eqnarray}
and Minkowski coordinates. It should be clear by the context which one is used. We also use the metric convention $g_{\mu \nu} = (1,-1,-1,-1)$.

\section{Effective Theory} 
\label{sec:eff_th}

The effective theory used in this article couples gluons and the $\eta'$ meson. In momentum space, the $g^{*}g^{*}\eta'$ effective vertex (where $g^{*}$ means off-shell gluon) is given by
\begin{eqnarray}
\label{eq:vertex}
V^{\mu \nu} (M,p,q) = i F(p^{2},q^{2},M^2) \delta_{ab} \epsilon^{\mu \nu \rho \alpha} p_{\rho} q_{\alpha}
\end{eqnarray}
where $\epsilon^{\mu \nu \rho \alpha}$ is the Levi-Civitta antisymmetric tensor, $M$ is the $\eta'$ mass, $a$ and $b$ are color indices, $p_{\rho}$ and $q_{\alpha}$ are gluon momenta and $F(p^{2},q^{2},M^2)$ is the $\eta'$ form factor. The explicit expression of the interaction vertex have been studied in a number of articles where different parametrizations of the form factor can be found \cite{Atwood:1997bn,Ali:2003kg,Ali:2000ci,Muta:1999tc,Ahmady:1998mi,Agaev:2002ek,Kroll:2002nt}. To get a first approximation of $\eta'$ production and because we are mostly interested in making a comparative study between $pp$ and $pA$ collisions, we use a simple expression given by \cite{Szczurek:2006bn}
\begin{eqnarray}
\label{eq:vertex2}
F(p^{2},q^{2},M^2)=H_{0} \frac{M^{4}}{(M^{2}-p^{2})(M^{2}-q^{2})}
\end{eqnarray}
where $H_{0} = F(0,0,M^{2})$. To get a better approximations of $\eta'$ production, other parametrizations should be used.   In the limit of no gluon virtualities ($p^{2},q^{2} =0 $), the form factor is a constant that can be fixed by looking at the decay of $\psi \rightarrow \eta'+\gamma$. It is  given by $H_{0} = F(0,0,M^2) \approx 1.8 \;\mbox{GeV}$ \cite{Atwood:1997bn}. 
 
The main applications of this coupling are related mostly to $B$ and $\Upsilon$ decay where processes such as $g^{*} \rightarrow g + \eta'$ and $g^{*} + g^{*} \rightarrow  \eta'$ are considered \cite{Atwood:1997bn,Kagan:1997qn,Hou:1997wy,Ahmady:1997fa,Du:1997hs}. More recently, gluon fusion was used to compute $\eta'$ production in high energy hadronic collisions \cite{Szczurek:2006bn,JalilianMarian:2001bu,Jeon:2002hh} and from a thermalized medium \cite{Jeon:2001cy}.

It is convenient for our purpose to consider the interaction Lagrangian given by 
\begin{eqnarray}
\label{eq:lag_int}
\mathcal{L}_{\rm{int}}(x) = \frac{1}{8} \int d^{4}y d^{4}z F\left[(x-y)^{2},(x-z)^{2},M^{2} \right] G_{a}^{\mu \nu}(y) \widetilde{G}_{a, \mu \nu}(z) \eta(x).
\end{eqnarray} 
that reproduces the vertex Eq. (\ref{eq:vertex}) in the perturbative expansion. As seen in the next section, this can then be used to derive a reduction formula. Here, $G_{a}^{\mu \nu}(x)$ is the usual field-strength tensor given by
\begin{eqnarray}
\label{eq:field_st}
G_{a}^{\mu \nu}(x) = \partial^{\mu} A_{a}^{\nu}(x) - \partial^{\nu} A_{a}^{\mu}(x) - gf_{abc}A_{b}^{\mu}(x)A_{c}^{\nu}(x)
\end{eqnarray}
where $A^{\mu}_{a}$ is the gauge field of gluons and $\tilde{G}_{a}^{\mu \nu}(x)=\epsilon^{\mu \nu \rho \sigma}G_{a,\rho \sigma}(x)$ is the dual field-strength tensor. The Lagrangian is non-local because the vertex includes a form factor. It can be easily seen that $\mathcal{L}_{\rm{int}}(x)$ contains three types of vertices, namely $gg\eta'$, $ggg\eta'$ and $gggg\eta'$. At leading order however, only the first one is necessary and considered in this article.


\section{Production of $\eta'$ from the CGC}
\label{sec:prod_CGC}

In collisions at very high energy, the wave function of nuclei is dominated by gluons that have small longitudinal momenta (soft gluons) because of the emission enhancement at small-$x$. The CGC is a semi-classical formalism that describes the dynamics of these degrees of freedom. In this approach, the hard partons, which carry most of the longitudinal momentum, and soft gluons which have small longitudinal components, are treated differently. Because the occupation number of the soft gluons is large, classical field equations can be employed to understand their dynamics. The hard partons act as sources for these classical field and are no longer interacting with the rest of the system (for reviews of CGC, see \cite{Iancu:2002xk,Iancu:2003xm,Venugopalan:2004dj}). 

In this formalism, computing a physical quantity involves two main steps. The first one is to solve the Yang-Mills equation of motion
\begin{eqnarray}
[D_{\mu},F^{\mu \nu}(x)] = J^{\nu}(x)
\end{eqnarray} 
where the current $J^{\nu}_{a}(x) = \delta^{\nu +} \delta(x^{-})\rho_{p,a}(x_{\perp}) + \delta^{\nu -} \delta(x^{+})\rho_{A,a}(x_{\perp})$ represents random static sources localized on the light-cone \cite{Iancu:2002xk,Iancu:2003xm} and $D^{\mu}=\partial^{\mu} - igA^{\mu}$ is the covariant derivative. The functions $\rho_{p,A}(x_{\perp})$ are color charge densities in the transverse plane of the proton and nucleus respectively. The next step is to take the average over the distribution of color charge densities in the nuclei with weight functionals $W_{p,A}[\rho_{p,A}]$. For any operator that can be related to color charge densities, this can be written as
\begin{eqnarray}
\label{eq:average}
\langle \hat{O} \rangle = \int \mathcal{D}\rho_{p} \mathcal{D} \rho_{A}  O[\rho_{p},\rho_{A}]W_{p}[x_{p},\rho_{p}]W_{A}[x_{A},\rho_{A}].
\end{eqnarray} 
Computing the weight functional is a highly non-perturbative procedure so it usually involves approximations based on physical modelling. In the limit of a large nuclei at not too small $x$, it can be approximated by the McLerran-Venugopalan (MV) model, which assumes that the partons are independent sources of color charge \cite{McLerran:1993ka,McLerran:1993ni}. Within this assumption, the weight functional $W_{A}[\rho_{A}]$ is a  $x_{A}$ independent Gaussian distribution and the two point correlator is simply \cite{McLerran:1993ka,McLerran:1993ni,Iancu:2002xk,Iancu:2003xm}
\begin{eqnarray}
\label{eq:MV1}
\langle \rho_{A,a}(x_{\perp}) \rho_{A,b}(y_{\perp}) \rangle = \delta_{ab}\mu_{A}^{2} \delta^{2}(x_{\perp} - y_{\perp})
\end{eqnarray}
where $\mu_{A}^{2}= A/2\pi R^{2}$ is the average color charge density and $R$ is the radius of the nucleus. It is assumed here that the nucleus has an infinite transverse extent with a constant charge distribution. Edge effects can be included by changing $\mu^{2} \rightarrow \mu^{2}(x_{\perp})$ and by choosing a suitable transverse profile. Throughout this article, we only consider the constant distribution case.

Within the MV model, the weight functional does not depend on longitudinal coordinates and therefore, the model is boost invariant. This however can be relaxed by considering the quantum version of the CGC. In that theory, quantum radiative corrections become important below a certain scale $x_{0} \approx 0.01$. These corrections can be resummed by using a renormalization group technique which leads to the JIMWLK evolution equation \cite{Iancu:2000hn,Ferreiro:2001qy,JalilianMarian:1996xn,JalilianMarian:1997gr,JalilianMarian:1997jx}. In the quantum CGC, the weight functionals $W_{1,2}[\rho_{1,2}]$ obey this non-linear evolution equation in $x$. Because the MV model is valid in the range $x \approx 0.01-0.1$, it can be used as an initial condition for the evolution at smaller $x$. In this article however, we consider only the regime where the MV model is valid and do not consider the small-$x$ evolution although it could be done in principle. 

On the proton side, the average computed with $W_{p}[x_{p},\rho_{p}]$ can be related to the unintegrated parton distribution function (uPDF) $\phi_{1}$ like 
\begin{eqnarray}
\label{eq:coor_unint_p}
 g^{2} \langle \rho_{p,a}^{*}(p_{\perp}) \rho_{p,b}(q_{\perp}) \rangle &=&\frac{4\pi^{2} \delta_{ab}}{ (N^{2}_{c}-1)}\left[\frac{p_{\perp}+q_{\perp}}{2} \right]^{2} \nonumber \\
&& \times \int d^{2}y_{\perp} e^{i(p_{\perp}-q_{\perp})\cdot y_{\perp}} \frac{d\phi_{1} \left(x,\frac{p_{\perp}+q_{\perp}}{2} | y_{\perp} \right)}{d^{2}y_{\perp}} 
\end{eqnarray} 
where $N_{c}$ is the number of color. By construction, the uPDF obeys
\begin{eqnarray}
\int d^{2}y_{\perp} \frac{d\phi_{1} \left(x,p_{\perp} | y_{\perp} \right)}{d^{2}y_{\perp}} = \phi_{1} (x,p_{\perp})
\end{eqnarray}
and is normalized such that
\begin{eqnarray}
\int_{0}^{\mu^{2}} \phi_{1,2} (x,p_{\perp}) \approx xG(x,\mu^{2})
\end{eqnarray}
where $xG(x,\mu^{2})$ is the collinear parton distribution function and $\mu^{2}$ is the factorization scale. The uPDF can be obtained from a fit to structure function and evolved to the desired value of $x_{p}$, $Q^{2}$ and $p_{\perp}^{2}$ using evolution equations such as the BFKL or the CCFM equations.

One important ingredient is missing for the computation of $\eta'$ meson production cross section. We need a relation between the cross section and a correlator that can be evaluated using the CGC formalism. This is done in the next section using a reduction formula and the effective theory.

\subsection{Reduction Formula and the Cross Section}
\label{sec:reduction}

The computation of $\eta'$ mesons from the CGC can be calculated from a reduction formula. The starting point is the expression of the average number of $\eta'$ produced per collisions given by $\bar{n} = \sum_{n=1}^{\infty} nP_{n}$ where $P_{n}$ is the probability to produce $n$ particles. This can be converted to an equation in terms of creation/annihilation operators that can be evaluated in quantum field theory. This is given by \cite{Baltz:2001dp}
\begin{eqnarray}
(2\pi)^3 2 E_k \frac{d\bar{n}}{d^3 k}
=
\bra{0_{\rm in}} 
\hat{a}^{\dagger}_{\rm out}(k) 
\hat{a}_{\rm out}(k) 
\ket{0_{\rm in}}
\label{eq:reduction_N_f}
\end{eqnarray}
where $\ket{0_{\rm in}}$ is the \textit{in} vacuum. Then, the standard LSZ procedure can be used to write this as \cite{Itzykson:1980rh} 
\begin{eqnarray}
(2\pi)^3 2 E_k \frac{d\bar{n}}{d^3 k}
= \frac{1}{\mathcal{Z}} \int d^{4}x d^{4}y e^{ik\cdot(x-y)} \left[\partial_{x}^{2} + M^2 \right] \left[\partial_{y}^{2} + M^2 \right]
\bra{0_{\rm in}} 
\hat{\eta}(x) \hat{\eta}(y)\ket{0_{\rm in}}
\end{eqnarray}
where $\mathcal{Z}$ is the wave function normalization and where we assumed the asymptotic conditions $\lim_{t \rightarrow \pm \infty}\hat{\eta}(x) = \sqrt{\mathcal{Z}} \hat{\eta}_{\rm{out,in}}(x)$ for the $\eta'$ field operator in Heisenberg representation. It is possible to use the equation of motion of $\hat{\eta}(x)$ given simply by 
\begin{eqnarray}
(\partial^2 + M^2) \hat{\eta} (x) &=& \hat{T}(x)
\label{eq:eom_H}
\end{eqnarray}
where we defined $\hat{T}(x) \equiv \frac{1}{8} \int d^{4}y d^{4}z F\left[(x-y)^{2},(x-z)^{2},M^{2} \right] \hat{G}_{a}^{\mu \nu}(y) \hat{\widetilde{G}}_{a, \mu \nu}(z)$ to rewrite the reduction formula in a convenient way. We get finally that
\be
(2\pi)^3 2E_k \frac{d\bar{n}}{d^3 k}
=
\langle
\hat{T}^{\dagger}(k)
\hat{T}(k)
\rangle
\label{eq:pre_main}
\ee
where $\hat{T}(k)$ is the Fourier transform of $\hat{T}(x)$ evaluated at a $\eta'$ meson on-shell momentum and where we set $\mathcal{Z}=1$. The angular brackets $\langle \hat{O} \rangle $ here indicates expectation value of $\hat{O}$ in the initial state. 

The only assumptions used in deriving Eq. (\ref{eq:pre_main}) are:
\begin{itemize}
	\item There are no $\eta'$ mesons in the initial state.
	\item The $\eta'$ is produced on-shell.
\end{itemize}
The first assumption is justified by the fact that in high-energy collisions, the number of $\eta'$ in a hadron before the collision (in the initial state) is negligible. This allows us to use the \textit{in} vacuum and the fact that $a_{\rm in}(k)|0_{\rm in} \rangle = 0$ to simplify the reduction formula. Using the second assumption, we can treat the $\eta'$ meson as a stable particle which can be produced on-shell and which is well described by the free spectral density that looks like $\rho(M^{2}) \sim \delta(p^{2}-M^{2})$. Therefore, by making this assumption, it is possible to use the asymptotic conditions described earlier. However, $\eta'$-mesons are resonances, so the spectral density should look rather as a Breit-Wigner function $\rho(M^{2}) \sim \Gamma/[(p^{2} - M^{2})^{2} + M^{2}\Gamma^{2}]$ where $\Gamma$ is the decay width. These effects however are taken into account by the form factor $F(q,p,M)$.  

Eq. (\ref{eq:pre_main}) is the main result of this section. It relates the average number of $\eta'$ mesons produced to a correlator of field strength tensors. This correlator can then be evaluated using any analytical or numerical methods. The average $\langle ... \rangle$ depends on the system studied. Looking at a plasma at equilibrium, it could be computed using finite temperature field theory or the AdS/CFT correspondence. These two formalisms are relevant to nucleus-nucleus collisions where a medium at equilibrium is created. We are interested here in $pA$ and $pp$ collisions where no such medium is formed so these techniques are not pursued in this study. Rather, we use the CGC which describes initial state and saturation effects in high-energy hadronic collisions.

Having expressed the average number of $\eta'$ produced in terms of a correlator of field strength tensors, it is possible to compute the inclusive cross section in the CGC formalism which is given by \cite{Gelis:2003vh,Baltz:2001dp}
\be
(2\pi)^3 2E_k \frac{d\sigma}{d^3 k}&=& \int d^{2}b_{\perp}(2\pi)^3 2E_k \frac{d\bar{n}(b_{\perp})}{d^3 k} \nonumber \\
&=&
 \int d^{2}b_{\perp}
\int \mathcal{D} \rho_{p} \mathcal{D} \rho_{A}
T^{*} [\rho_{p},\rho_{A}]
T[\rho_{p},\rho_{A}] \nonumber \\
&& \times
W_{p}[x_{p},\rho_{p}]W_{A}[\rho_{A};b_{\perp}].
\label{eq:cross_CGC_1}
\ee
In this expression, $b_{\perp}$ is the impact parameter. The fields $T$ are functionals of the source once the Yang-Mills equation of motion of the gauge field is solved (see Eq. (\ref{eq:field_st}) for the expression of the field strength tensor as a function of the gauge field).

\subsection{Cross section in $pA$ Collisions}

In $pA$ collisions, there are two saturation scales (one for the proton ($Q_{p}$) and one for the nucleus ($Q_{A}$)) that satisfy $Q_{p}<Q_{A}$. When the transverse momentum of the $\eta'$ is small enough, the nucleus is in a saturation state while the proton is not because we have $Q_{p}^{2} < Q^{2} = M_{\perp}^{2} < Q_{A}^{2}$ (remember that $M_{\perp}^{2} = M^{2}+k_{\perp}^{2}$ is the transverse mass of the $\eta'$). In that case, the system is semi-dilute, meaning that one of the source is strong (or equivalently, the typical transverse momentum is small) and obeys $\rho_{A,a}(k_{\perp}) /k_{\perp}^{2} \sim 1$ while the other source is still weak $\rho_{p,a}(k_{\perp}) /k_{\perp}^{2} \ll 1$ \cite{Venugopalan:2004dj,Gelis:2004bz}. The weak source can be used as a small parameter to solve the Yang-Mills equation perturbatively. The solution of the gauge field can be computed analytically to all orders in  $\rho_{A,a}(k_{\perp}) /k_{\perp}^{2}$ and to first order in $\rho_{p,a}(k_{\perp}) /k_{\perp}^{2}$ in different gauges \cite{Blaizot:2004wu,Dumitru:2001ux,Gelis:2005pt,Kovchegov:1998bi}. We use here the solution in the light-cone gauge of the proton \cite{Gelis:2005pt} but in Appendix \ref{app:cov}, we perform the same calculation in covariant gauge to show that our result is gauge invariant.

\subsubsection{Gauge Field and Power Counting}

In the light-cone gauge of the proton ($A^{+} = 0$) with a nucleus in covariant gauge moving in the negative $z$ direction, the solution of the gauge field in $pA$ collisions can be separated in three parts $A^{\mu}_{a}(k) = A^{\mu}_{p , a}(k) + A^{\mu}_{A,a}(k) + A^{\mu}_{pA,a}$ where $A^{\mu}_{p,a}(k)$ is the field associated with the proton (of $O(\rho_{p})$), $A^{\mu}_{A, a}(k)$ is the field associated with the nuclei (of $O(\rho_{A}) \sim O(1)$) and $A^{\mu}_{pA,a}(k)$ is the field produced by the collision (of $O(\rho_{p}\rho_{A}^{\infty}) \sim O(\rho_{p})$) \cite{Gelis:2005pt}. Note that the field $A^{\mu}_{A}$ is strong and satisfies $A^{\mu}_{p},A^{\mu}_{pA} \ll A^{\mu}_{A}$. The explicit solution is given by \cite{Gelis:2005pt,Fukushima:2008ya}
\begin{eqnarray}
\label{eq:gauge_p}
A^{i}_{p ,a}(k) &=& 2\pi g  \delta (k^{-})\frac{k^{i}}{k^{+}+i \epsilon} \frac{\rho_{p, a} (k_{\perp})}{k_{\perp}^{2}}  \\
\label{eq:gauge_A}
A^{-}_{A,a}(k) &=& 2\pi g \delta(k^{+}) \frac{\rho_{A,a}(k_{\perp})}{k_{\perp}^{2}} \\
\label{eq:gauge_prod}
A^{i}_{pA,a}(k) &=& -\frac{ig}{k^{2} + ik^{+} \epsilon} \int \frac{d^{2}p_{\perp}}{(2\pi)^{2}} \left[ \frac{k^{i}}{(k^{+} + i \epsilon)(k^{-} + i \epsilon)} - 2\frac{p^{i}}{p_{\perp}^{2}} \right] \rho_{p,b}(p_{\perp}) \nonumber \\
&& \times \left[ U_{ab}(k_{\perp} - p_{\perp}) - (2\pi)^{2} \delta^{2}(k_{\perp} - p_{\perp}) \delta_{ab}   \right]
\end{eqnarray} 
where $U_{ab}(k_{\perp})$ is a Wilson line in adjoint representation defined in Eq. (\ref{eq:wilson}), $g$ is the usual QCD coupling constant, $\delta_{ab}$ is the Kronecker delta in color space and $f_{abc}$ is the structure constant of the $SU(N_{c})$ group. The component $A^{-}_{pA,a}(k)$ is non-zero and is related to $A^{i}_{pA,a}(k)$ but it does not appear in the final expression of the cross section so it is not needed in the computation of $\eta'$ production. All the other components are zero.

The production cross section of $\eta'$ mesons is related to a field strength tensor correlator given by
\begin{eqnarray}
 B(k) & \equiv & \int \frac{d^{4}p d^{4} q}{(2\pi)^{8}} F(p^{2} , p_{2}^{2})  F^{*}(q^{2},q_{2}^{2})  \langle G_{\mu \nu a}^{*}(p)\widetilde{G}^{* \mu \nu}_{a}(k-p) 
G_{\alpha \beta b}(q)\widetilde{G}^{\alpha \beta}_{b}(k-q) \rangle 
\end{eqnarray}
where $p_{2},q_{2}=k-p,q$. In principle, a correlator like this contains contributions from all orders in both sources. Because the proton source is weak, it is possible to simplify this considerably using a power counting argument to isolate the leading order contribution. For the sake of this power counting argument, we use $a_{(...)}^{n}$ which denotes terms having $n$ gauge fields $A_{(...)}^{\mu}$ and where $a_{A} \sim O(\rho_{A})$ and $a_{p},a_{pA} \sim O(\rho_{p})$. At first, let us consider only the contributions from the abelian part of the field-strength tensor. The terms in these contributions have four powers of gauge field such as $B_{\rm{abelian}} \sim a_{\rm{tot}}^{4} \sim (a_{A} + a_{p}+ a_{pA})^{4}$. Naively, one would expect the dominant contribution to come from terms that have many powers of the nucleus gauge field like $a_{A}^{4} \sim O(\rho^{4}_{A})$ and  $a_{A}^{3}a_{p},a_{A}^{3}a_{pA} \sim O(\rho^{3}_{A} \rho_{p})$. However, these terms vanish because of the Lorentz structure. For example, a typical term in the abelian contribution would look like $T_{\mathrm{abelian}} \sim \epsilon_{\mu \nu \rho \sigma}p^{\mu} q^{\nu} A^{\rho} A^{\sigma}$. When we sum on indices, this kind of term will contain at most one strong gauge field $A^{-} \sim a_{A}$. Thus, the dominant contributions are like $a_{A}^{2}a_{p}^{2},a_{A}^{2}a_{pA}^{2} \sim O(\rho^{2}_{A} \rho^{2}_{p})$. Using a similar argument, it is possible to show that the non-abelian part have no leading order contribution in the sense that it is at least $B_{\rm{non-abelian}} \sim O(\rho^{2}_{A} \rho^{3}_{p}) \ll O(\rho^{2}_{A} \rho^{2}_{p})$. The possible higher order contributions like $a_{A}^{3}a_{p}^{2} \sim O(\rho^{3}_{A} \rho^{2}_{p})$ for example also vanish because of the Lorentz structure of the correlator. This is because the typical non-abelian contributions look like $T'_{\mathrm{non-abelian}} \sim \epsilon_{\mu \nu \rho \sigma}p^{\mu} A^{\nu} A^{\rho} A^{\sigma}$ and $T''_{\mathrm{non-abelian}} \sim \epsilon_{\mu \nu \rho \sigma}A^{\mu} A^{\nu} A^{\rho} A^{\sigma}$. Once the Lorentz indices are summed, the second typical term $T''=0$ because in this gauge, $A^{+}=0$. For $T'$, it contains only one strong field $A^{-} \sim a_{pA}$ but it contains two powers of weak field like $A^{i} \sim a_{p},a_{pA}$. Thus, when it is squared, it gives at most a contribution of $B_{\rm{non-abelian}} \sim O(\rho^{2}_{A} \rho^{3}_{p})$.

\subsection{Evaluation of the Correlator}
\label{sec:calc_corr}

Using the explicit expression of the field strength tensor in terms of gauge field and keeping only the dominant and non-zero contributions, the correlator can be written as 
\begin{eqnarray}
\label{eq:corr_G}
B(k) &=& 64  \int \frac{d^{4}p d^{4} q}{(2\pi)^{8}} F(-p_{\perp}^{2} , -p_{2,\perp}^{2})  F^{*}(-q_{\perp}^{2},-q_{2,\perp}^{2})  (k^{+} - p^{+})(k^{+} - q^{+}) \epsilon_{ij}\epsilon_{kl}p^{i} q^{k} \nonumber \\
&& \times  \langle A_{A,a}^{-*} (p) A_{A,b}^{-} (q) A_{s,a}^{j *}(k-p)  A_{s,b}^{l }(k-q)   \rangle 
\end{eqnarray}
where $\epsilon_{ij}$ is the Levi-Civitta antisymmetric tensor with $i,j=1,2$ and where we defined $A_{s,a} \equiv A_{p,a} + A_{pA,a}  $.

To obtain the preceding expression we make the assumption that the virtualities in the form factors are due solely to transverse momentum such as $F(p^{2} , p_{2}^{2}) = F(-p_{\perp}^{2} , -p_{2,\perp}^{2})$. This approximation is necessary to recover $k_{\perp}$-factorization in the dilute limit as seen in section \ref{sec:k_perp}.

It is convenient to separate $B(k)$ in four different terms such as 
\begin{eqnarray}
B_{z,z'}(k) &=& 64g^{2} (k^{+})^{2} \int \frac{dp^{-} dq^{-}d^{2}p_{\perp} d^{2} q_{\perp}}{(2\pi)^{6}}  F(-p_{\perp}^{2} , -p_{2,\perp}^{2}) F^{*}(-q_{\perp}^{2},-q_{2,\perp}^{2})  \epsilon_{ij}\epsilon_{kl} \frac{p^{i} q^{k}}{p^{2}_{\perp} q^{2}_{\perp}}  \nonumber \\
&& \times \left. \langle \rho_{A,a}^{*} (p_{\perp}) \rho_{A,b} (q_{\perp}) A_{z,a}^{j *}(k-p)  A_{z',b}^{l }(k-q)   \rangle \right|_{p^{+} = q^{+}=0}  
\end{eqnarray} 
where $z,z'=\{p,pA\}$ and where we performed the integration on $p^{+}$ and $q^{+}$ using the delta functions in Eq. (\ref{eq:gauge_A}). These four terms can be evaluated explicitly by substituting the solution of gauge fields Eqs. (\ref{eq:gauge_p}), (\ref{eq:gauge_A}) and (\ref{eq:gauge_prod}).

For the first term, it is a straightforward calculation to show that
\begin{eqnarray}
B_{p,p}(k) &=& 64g^{4}  \int \frac{d^{2}p_{\perp} d^{2}q_{\perp} d^{2}r_{\perp} d^{2}s_{\perp}}{(2\pi)^{8}}F(-p_{\perp}^{2} , -p_{2,\perp}^{2}) F^{*}(-q_{\perp}^{2},-q_{2,\perp}^{2})\nonumber \\
&& \times  \epsilon_{ij}\epsilon_{kl} \frac{p^{i} r^{j} q^{k} s^{l}}{p^{2}_{\perp} q^{2}_{\perp} r^{2}_{\perp} s^{2}_{\perp}}   \delta_{bc} (2\pi)^{2} \delta^{2}(k_{\perp} - p_{\perp} - r_{\perp}) \delta_{ad} (2\pi)^{2} \delta^{2}(k_{\perp} - q_{\perp} - s_{\perp}) \nonumber \\
&& \times \langle \rho_{p,d}^{*}(r_{\perp}) \rho_{p,c}(s_{\perp}) \rangle \langle \rho_{A,a}^{*}(p_{\perp}) \rho_{A,b}(q_{\perp}) \rangle .   
\end{eqnarray}

The calculation of the second term is similar but requires some more work. By direct substitution of the expression of the gauge fields, we have 
\begin{eqnarray}
B_{pA,p}(k) &=& 64ig^{4} (k^{+}) \int \frac{dp^{-} d^{2}p_{\perp} d^{2} q_{\perp} d^{2}r_{\perp}}{(2\pi)^{7}}F(-p_{\perp}^{2} , -p_{2,\perp}^{2}) F^{*}(-q_{\perp}^{2},-q_{2,\perp}^{2}) \nonumber \\
&& \times  \epsilon_{ij}\epsilon_{kl} \frac{p^{i} q^{k} q_{2}^{l}}{p^{2}_{\perp} q^{2}_{\perp} q_{2,\perp}^{2}}     \frac{1}{p_{2}^{2} - ik^{+}\epsilon} \left[ \frac{p_{2}^{j}}{(p_{2}^{+} - i\epsilon)(p_{2}^{-} -  i\epsilon)}  -2 \frac{r^{j}}{r_{\perp}^{2}}  \right] \nonumber \\
&& \times \biggl[ \langle \rho_{A,a}^{*} (p_{\perp}) \rho_{A,b} (q_{\perp}) U^{*}_{ac} (p_{2,\perp} - r_{\perp}) \rangle \langle \rho_{p,c}^{*} (r_{\perp}) \rho_{p,b} (q_{2,\perp}) \rangle \nonumber \\
&& - (2\pi)^{2} \delta^{2}(p_{2,\perp} - r_{\perp})\delta_{ac} \langle \rho_{A,a}^{*} (p_{\perp}) \rho_{A,b} (q_{\perp})  \rangle\langle \rho_{p,c}^{*} (r_{\perp}) \rho_{p,b} (q_{2,\perp}) \rangle \biggr].
\end{eqnarray}
The integration on the longitudinal momentum $p^{-}$ can be done by looking at the analytical structure of the equation. First, we write a part of the integrand as
\begin{eqnarray}
\label{eq:integrand}
\left. \frac{1}{(k-p)^{2} - ik^{+}\epsilon}\left[ \frac{p_{2}^{j}}{(k^{+} - i\epsilon)(k^{-} - p^{-} -  i\epsilon)}  -2 \frac{r^{j}}{r_{\perp}^{2}}  \right]\right|_{p^{+}=0} = \nonumber \\
\frac{1}{2k^{+} \left[k^{-}-p^{-} - \frac{(k-p)_{\perp}^{2}}{2k^{+}} - i\epsilon \right] } \left[ \frac{p_{2}^{j}}{(k^{+} - i\epsilon)(k^{-} - p^{-} -  i\epsilon)}  -2 \frac{r^{j}}{r_{\perp}^{2}}  \right].
\end{eqnarray}
In the complex plane of $p^{-}$, the first term in the RHS of Eq. (\ref{eq:integrand}) has two poles on the same side of the real axis and goes like $\frac{1}{(p^{-})^{2}}$ when $p^{-} \rightarrow \infty$. Thus, closing the integration contour in the upper-half plane and using the residue theorem, the integration on $p^{-}$ of this term leads to a zero contribution because the contour at infinity has a zero contribution and because the contour does not enclose any singularities. For the second term of Eq. (\ref{eq:integrand}), we use the principal part ($\rm{PP}$) identity $\frac{1}{x\pm i\epsilon} = \rm{PP} \frac{1}{x} \mp i\pi \delta(x)$. The integration on the principal part is zero because the integrand does not depend on $p^{-}$ and $\lim_{R \rightarrow \infty} \int_{-R}^{R}dp \rm{PP} \frac{1}{p-a}=0$ while the delta function integration is trivial. We finally get 
\begin{eqnarray}
B_{pA,p}(k) &=&  32g^{4} \int \frac{d^{2}p_{\perp} d^{2}q_{\perp} d^{2}r_{\perp} d^{2}s_{\perp}}{(2\pi)^{8}} F(-p_{\perp}^{2} , -p_{2,\perp}^{2}) F^{*}(-q_{\perp}^{2},-q_{2,\perp}^{2})     \nonumber \\
&& \times \epsilon_{ij}\epsilon_{kl} \frac{p^{i} q^{k} r^{j} s^{l}}{p^{2}_{\perp} q^{2}_{\perp} r_{\perp}^{2} s_{\perp}^{2}} \langle \rho_{p,d}^{*}(r_{\perp}) \rho_{p,c}(s_{\perp}) \rangle \nonumber \\
&& \times \biggl[ -\langle \rho_{A,a}^{*}(p_{\perp}) \rho_{A,b}(q_{\perp}) \rangle \delta_{bc} (2\pi)^{2} \delta^{2}( p_{2,\perp} - r_{\perp}) \delta_{ad} (2\pi)^{2} \delta^{2}( q_{2,\perp} - s_{\perp}) \nonumber \\
&&+ \langle \rho_{A,a}^{*} (p_{\perp}) \rho_{A,b} (q_{\perp}) U^{*}_{ad} ( p_{2,\perp} - r_{\perp}) \rangle \delta_{bc} (2\pi)^{2} \delta^{2}( q_{2,\perp} - s_{\perp}) \biggr] .
\end{eqnarray}
The calculations of $B_{pA,p}(k)$ and $B_{pA,pA}(k)$ are similar. Going through the same steps as for $B_{p,pA}(k)$, we get
\begin{eqnarray}
B_{p,pA}(k) &=&  32g^{4} \int \frac{d^{2}p_{\perp} d^{2}q_{\perp} d^{2}r_{\perp} d^{2}s_{\perp}}{(2\pi)^{8}} F(-p_{\perp}^{2} , -p_{2,\perp}^{2}) F^{*}(-q_{\perp}^{2},-q_{2,\perp}^{2})  \nonumber \\
&& \times \epsilon_{ij}\epsilon_{kl} \frac{p^{i} q^{k} r^{j} s^{l}}{p^{2}_{\perp} q^{2}_{\perp} r_{\perp}^{2} s_{\perp}^{2}}  \langle \rho_{p,d}^{*}(r_{\perp}) \rho_{p,c}(s_{\perp}) \rangle \nonumber \\
&& \times \biggl[ -\langle \rho_{A,a}^{*}(p_{\perp}) \rho_{A,b}(q_{\perp}) \rangle \delta_{bc} (2\pi)^{2} \delta^{2}( p_{2,\perp} - r_{\perp}) \delta_{ad} (2\pi)^{2} \delta^{2}(q_{2,\perp} - s_{\perp}) \nonumber \\
&&+  \langle \rho_{A,a}^{*} (p_{\perp}) \rho_{A,b} (q_{\perp}) U_{bc} ( q_{2,\perp} - s_{\perp}) \rangle \delta_{ad} (2\pi)^{2} \delta^{2}( p_{2,\perp} - r_{\perp}) \biggr]
\end{eqnarray}
and
\begin{eqnarray}
B_{pA,pA}(k)&=&  16g^{4} \int \frac{d^{2}p_{\perp} d^{2}q_{\perp} d^{2}r_{\perp} d^{2}s_{\perp}}{(2\pi)^{8}} F(-p_{\perp}^{2} , -p_{2,\perp}^{2}) F^{*}(-q_{\perp}^{2},-q_{2,\perp}^{2}) \nonumber \\
&& \times \epsilon_{ij}\epsilon_{kl} \frac{p^{i} q^{k} r^{j} s^{l}}{p^{2}_{\perp} q^{2}_{\perp} r_{\perp}^{2} s_{\perp}^{2}}  \langle \rho_{p,d}^{*}(r_{\perp}) \rho_{p,c}(s_{\perp}) \rangle \nonumber \\
&& \times \biggl[ \langle \rho_{A,a}^{*}(p_{\perp}) \rho_{A,b}(q_{\perp}) \rangle \delta_{bc} (2\pi)^{2} \delta^{2}(p_{2,\perp} - r_{\perp}) \delta_{ad} (2\pi)^{2} \delta^{2}( q_{2,\perp} - s_{\perp}) \nonumber \\
&&-  \langle \rho_{A,a}^{*} (p_{\perp}) \rho_{A,b} (q_{\perp}) U_{bc} ( q_{2,\perp} - s_{\perp}) \rangle \delta_{ad} (2\pi)^{2} \delta^{2}( p_{2,\perp} - r_{\perp}) \nonumber \\
&&- \langle \rho_{A,a}^{*} (p_{\perp}) \rho_{A,b} (q_{\perp}) U^{*}_{ad} ( p_{2,\perp} - r_{\perp}) \rangle \delta_{bc} (2\pi)^{2} \delta^{2}( q_{2,\perp} - s_{\perp}) \nonumber \\
&&+  \langle \rho_{A,a}^{*} (p_{\perp}) \rho_{A,b} (q_{\perp})U^{*}_{ad} ( p_{2,\perp} - r_{\perp}) U_{bc} ( q_{2,\perp} - s_{\perp})  \rangle \biggr] .
\end{eqnarray}
The final result for the correlator $B(k)=B_{p,p}(k)+B_{p,pA}(k)+B_{pA,p}(k)+B_{pA,pA}(k)$ can be written compactly as
\begin{eqnarray}
B(k) &=& 16g^{4} \int \frac{d^{2}p_{\perp} d^{2} q_{\perp} d^{2}r_{\perp} d^{2}s_{\perp}}{(2\pi)^{8}}F(-p_{\perp}^{2} , -p_{2,\perp}^{2}) F^{*}(-q_{\perp}^{2},-q_{2,\perp}^{2}) \epsilon_{ij}\epsilon_{kl} \frac{p^{i} q^{k} r^{j} s^{l}}{p^{2}_{\perp} q^{2}_{\perp} r_{\perp}^{2} s_{\perp}^{2}} \nonumber \\
&& \times  \langle \rho_{p,d}^{*} (r_{\perp}) \rho_{p,c} (s_{\perp}) \rangle   \langle \rho_{A,a}^{*} (p_{\perp}) \rho_{A,b} (q_{\perp}) \tilde{U}^{*}_{ad} ( p_{2,\perp} - r_{\perp}) \tilde{U}_{bc} (q_{2,\perp} - s_{\perp}) \rangle 
\end{eqnarray}
where we defined
\begin{eqnarray}
\tilde{U}_{ab}(k_{\perp}) &=& U_{ab}(k_{\perp}) + (2\pi)^{2} \delta^{2}(k_{\perp} )  \delta_{ab}.
\end{eqnarray} 
From this equation for $B(k)$,  we can now evaluate the differential cross section. It is given by
\begin{eqnarray}
\label{eq:cross01}
(2\pi)^3 2E_k \frac{d\sigma}{d^3 k}
&=&
\frac{g^{4}}{4} \int d^{2}b_{\perp} \int \frac{d^{2}p_{\perp} d^{2} q_{\perp} d^{2}r_{\perp} d^{2}s_{\perp}}{(2\pi)^{8}}F(-p_{\perp}^{2} , -p_{2,\perp}^{2}) F^{*}(-q_{\perp}^{2},-q_{2,\perp}^{2}) \nonumber \\
&& \times \epsilon_{ij}\epsilon_{kl} \frac{p^{i} q^{k} r^{j} s^{l}}{p^{2}_{\perp} q^{2}_{\perp} r_{\perp}^{2} s_{\perp}^{2}}   \langle \rho_{p,c}^{*} (r_{\perp}) \rho_{p,d} (s_{\perp}) \rangle \nonumber \\
&& \times   \langle \rho_{A,a}^{*} (p_{\perp}) \rho_{A,b} (q_{\perp}) \tilde{U}^{*}_{ac} ( p_{2,\perp} - r_{\perp}) \tilde{U}_{bd} (q_{2,\perp} - s_{\perp}) \rangle .
\end{eqnarray} 
Thus, we can relate the $\eta'$ production cross section to a correlator of sources and Wilson lines. This will be evaluated in the next section in the MV model. Clearly, this expression does not have the $k_{\perp}$-factorization  structure but it can be recovered in the dilute limit. First, we can look at the physical interpretation of the different terms in the cross section.

\subsection{Diagrammatic Content and Physical Interpretation}

\begin{figure}
\centering
		\includegraphics{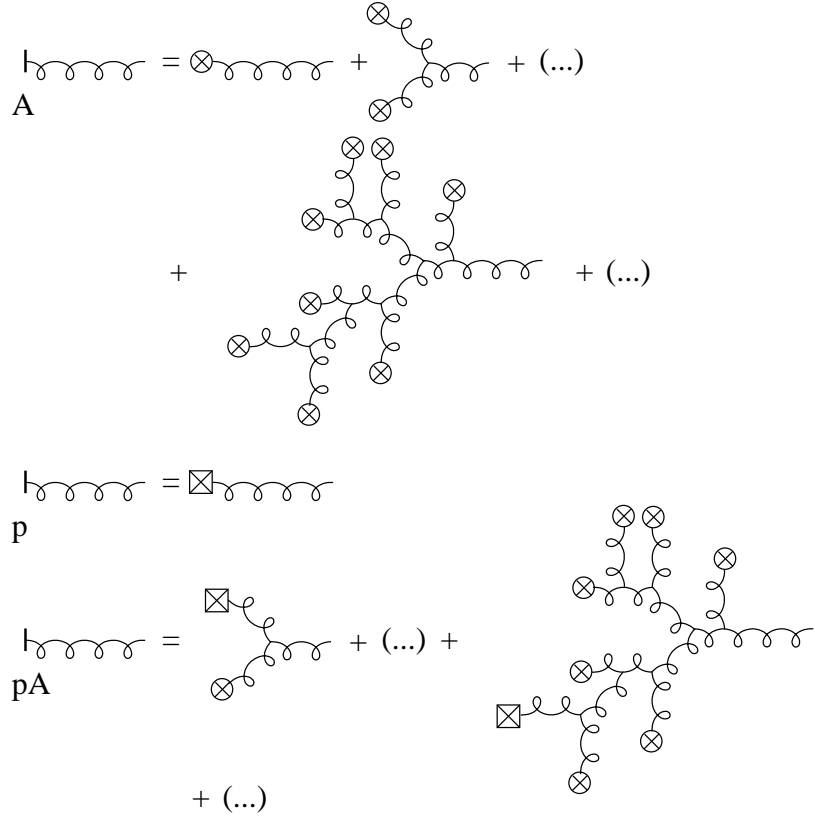}
	\caption{Diagrams included in the gauge field $A^{\mu}_{A}$,$A^{\mu}_{p}$ and $A^{\mu}_{pA}$. The crossed circles $\otimes$ represent insertions of the strong source $\rho_{A,a}(x_{\perp})$ while the crossed squares $\boxtimes$ represent insertions of the weak source $\rho_{p,a}(x_{\perp})$. By solving the Yang-Mills equation with retarded boundary conditions, it resums all the tree diagrams such as the ones depicted in this figure \cite{Gelis:2006yv}.}
	\label{fig:gauge_dia}
\end{figure}

To facilitate the physical interpretation, it is convenient to interpret the gauge field expressed in Eqs. (\ref{eq:gauge_p}),(\ref{eq:gauge_A}) and (\ref{eq:gauge_prod}) in terms of Feynman diagrams as shown in Fig. \ref{fig:gauge_dia}. The strong gauge field $A^{\mu}_{A}$ corresponds to a resummation of tree diagrams with any number of strong source insertions \cite{Gelis:2006yv}. The weak gauge fields $A^{\mu}_{p}$ and $A^{\mu}_{pA}$ also resum an infinite number of tree diagrams, the difference being that they contain one weak source insertion \cite{Blaizot:2004wu}.

\begin{figure}
\centering
		\includegraphics{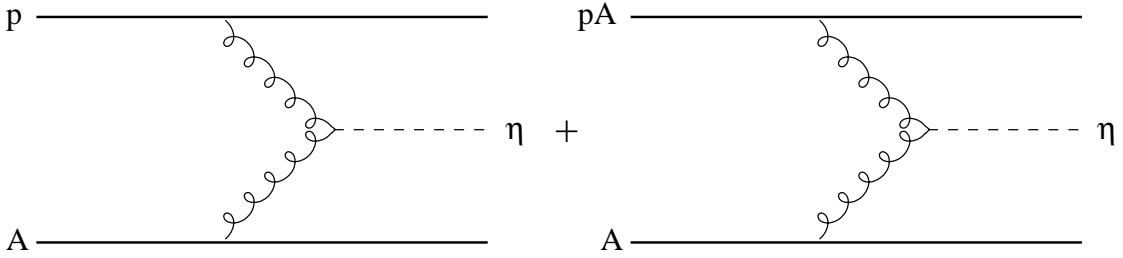}
	\caption{Diagrams included in $\eta'$ production at leading order. The thick lines represent insertions of the proton ($p$), nucleus ($A$) and produced field ($pA$),  and the dashed line is the $\eta'$ meson. The field $A^{\mu}_{pA}$ contains multi-scattering diagrams shown in Fig. \ref{fig:higgs_dia2}. The first figure represents the interaction between two gluons producing a $\eta'$ meson that goes through the nucleus without interacting. The second figure corresponds to the situation where the gluons emitted by the proton goes through the nucleus before producing the $\eta'$. }
	\label{fig:higgs_dia}
\end{figure}

\begin{figure}
\centering
		\includegraphics{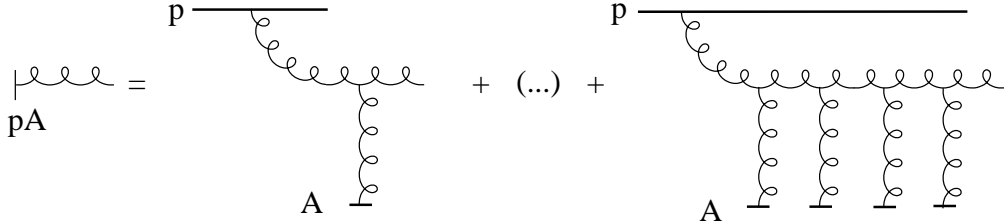}
	\caption{Multi-scattering diagrams included in $A^{\mu}_{pA}$ shown in Fig \ref{fig:higgs_dia}. The last diagram is a typical diagram associated with the Wilson line. This field contains the multiscattering effects.}
	\label{fig:higgs_dia2}
\end{figure}

Then, the $\eta'$ production and the correlator given in Eq. (\ref{eq:corr_G}) can be represented diagrammatically in Fig. \ref{fig:higgs_dia} and \ref{fig:higgs_dia2}. These figures show all the diagrams included in the calculation. The first term in the figure corresponds to the part where the $\eta'$ is produced from gluon fusion. The second term contains multiscattering effects and eventually, saturation effects. Overall, this leads to the following physical picture. A gluon inside the proton interact with the classical background field of the nucleus and gets multiscattered. Once it has gone through the nucleus, it combines with a gluon and produce the $\eta'$.


\subsection{Recovering $k_{\perp}$-factorization in the cross section}
\label{sec:k_perp}

It is possible to recover $k_{\perp}$-factorization from Eq. (\ref{eq:cross01}) by looking at the dilute limit of the nucleus characterized by a weak source such as $\rho_{A,a} \ll 1$. In that case, we are allowed to keep only the first term of the Wilson line expansion $\tilde{U}_{ae}(k_{\perp}) = 2(2\pi)^{2} \delta^{2}(k_{\perp}) \delta_{ae} + O(\rho_{A})$. In this low-density limit, the cross section becomes
\begin{eqnarray}
(2\pi)^3 2E_k \frac{d\sigma_{\rm{low-density}}}{d^3 k}
&=&
g^{4} \int d^{2}b_{\perp} \int \frac{d^{2}p_{\perp} d^{2} q_{\perp} }{(2\pi)^{4}}F(-p_{\perp}^{2} , -p_{2,\perp}^{2}) F^{*}(-q_{\perp}^{2},-q_{2,\perp}^{2}) \nonumber \\
&& \times \epsilon_{ij}\epsilon_{kl} \frac{p^{i} q^{k} p_{2}^{j} q_{2}^{l}}{p^{2}_{\perp} q^{2}_{\perp} p_{2,\perp}^{2} q_{2,\perp}^{2}}   \langle \rho_{p,a}^{*} (p_{2,\perp}) \rho_{p,b} (q_{2,\perp}) \rangle \langle \rho_{A,a}^{*} (p_{\perp}) \rho_{A,b} (q_{\perp})  \rangle .
\end{eqnarray}
In terms of Feynman diagrams, this expression corresponds to neglecting all the multi-scattering diagrams shown in Fig. \ref{fig:higgs_dia2}. The neglected diagrams are the ones that break $k_{\perp}$-factorization as can be seen from the following argument.

In this dilute limit, the correlator of the nucleus can also be related to the uPDF $\phi_{2}$ like in Eq. (\ref{eq:coor_unint_p}). Using this results, we find that the cross section is given by
\begin{eqnarray}
\label{eq:cross_kfac}
(2\pi)^3 2E_k \frac{d\sigma_{\rm{low-density}}}{d^3 k}
&=&
\frac{ 4 \pi^{2}}{ (N_{c}^{2}-1)} 
\int d^{2}p_{\perp}  d^{2}q_{\perp}|F(-p_{\perp}^{2} , -q_{\perp}^{2})|^{2}  \nonumber \\
&& \times \epsilon_{ij}\epsilon_{kl} \frac{p^{i} p^{k} q^{j} q^{l}}{p^{2}_{\perp} q^{2}_{\perp}}    \phi_{1}(q_{\perp}) \phi_{2}(p_{\perp})   \delta^{2} (k_{\perp}-p_{\perp}-q_{\perp}).  
\end{eqnarray}
This is the $k_{\perp}$-factorized expression of the cross section and is totally equivalent to Eq. (\ref{eq:k_fact_cross}) obtained directly from the $k_{\perp}$-factorization formalism (see Appendix \ref{app:pp_k_perp}). Thus, in the low-density limit of $pA$ collisions, we recover a formalism that describes $pp$ collisions in the semihard regime. This is very similar to quark and gluon production in $pA$ collisions \cite{Gelis:2003vh,Gyulassy:1997vt,Blaizot:2004wu,Blaizot:2004wv}. Note that to obtain this result, it is necessary to assume from the beginning that the form factors depend only on transverse momenta.

\section{Computation of Correlation Functions in the MV Model}
\label{sec:averages}

In this section, we compute the relevant correlators appearing in our expression of the cross section using the McLerran-Venugopalan (MV) model. Throughout this calculation, we use the notation of \cite{Gelis:2001da,Fukushima:2007dy}. We are interested in correlators containing both Wilson lines and color charge densities such as the ones included in Eq. (\ref{eq:cross01}). In App. \ref{app:dia_rules}, we discuss the general case and give more details on the calculation. Note here that to make sense of the ordered path, we start with color charge densities that depend on the longitudinal coordinate $x^{+}$. At the end of the calculation, we take $\rho_{a}(x^{+},x_{\perp}) = \delta(x^{+})\rho_{a}(x_{\perp})$ since we use the MV model which assumes that the nucleus is moving at the speed of light. In this more general case, the 2-point function is simply
\begin{eqnarray}
\label{eq:MV}
\langle \rho_{a}(x^{+},x_{\perp}) \rho_{b}(y^{+},y_{\perp}) \rangle = \delta_{ab}\mu^{2}(x^{+})\delta(x^{+} - y^{+}) \delta^{2}(x_{\perp} - y_{\perp})
\end{eqnarray}
where $\mu^{2}(x^{+})$ is the average color charge density at point $x^{+}$. It is related to the average color charge density by $\mu^{2} = \int dx^{+} \mu^{2}(x^{+}) = A/2\pi R^{2}$ where $R$ is the radius of the nucleus. In this model, $W[\rho]$ is still Gaussian, so all even-point functions can be written in terms of the 2-point function using Wick theorem and all odd-point functions are zero.

The Wilson line is defined as 
\begin{eqnarray}
\label{eq:wilson}
U_{ab}(b^{+},a^{+}|x_{\perp}) = \mathcal{P}^{+} \exp \left[ -ig^{2} \int_{a^{+}}^{b^{+}} dz^{+}\int d^{2}z_{\perp} G_{0}(x_{\perp} - z_{\perp}) \rho_{c}(z^{+},z_{\perp})t^{c}  \right]_{ab}
\end{eqnarray}
where $t^{c}$ are the $SU(N_{c})$ generators in adjoint representation, $\mathcal{P}^{+}$ is the path ordering in the light-cone coordinate $z^{+}$ and $G_{0}$ is a Green function solution of
\begin{eqnarray}
\frac{\partial^{2}}{\partial x^{2}_{\perp}}G_{0}(x_{\perp}) = \delta^{2}(x_{\perp}).
\end{eqnarray}

With these definitions, it is possible to compute the needed correlators. The general strategy is to express the correlators in terms of the following known results for Wilson lines in adjoint representation \cite{Fukushima:2007dy}:
\begin{eqnarray}
\label{eq:one_av}
\langle U_{ab}(b^{+},a^{+}|x_{\perp}) \rangle &\equiv & \bar{U}(b^{+},a^{+}|x_{\perp}) \delta_{ab} \nonumber \\
&=& \delta_{ab} \exp \left[ -\frac{N_{c}}{2} L(x,x) \bar{\mu}^{2}(b^{+},a^{+})  \right] \\
\langle U_{ab}(b^{+},a^{+}|x_{1\perp}) U_{cd}(b^{+},a^{+}|x_{2\perp})  \rangle & \equiv & \frac{\delta_{ac}
\label{eq:two_av}
\delta_{bd}}{N_{c}^{2}-1} \bar{V}(b^{+},a^{+}|x_{1\perp},x_{2\perp}) \nonumber \\
&=& \frac{\delta_{ac} \delta_{bd}}{N_{c}^{2}-1} \exp \left[- N_{c} \bar{\mu}^{2}(b^{+},a^{+}) \left( L(0,0) - L(x_{1 \perp},x_{2 \perp}) \right)    \right]
\end{eqnarray}
where
\begin{eqnarray}
\label{eq:L_00}
L(x,y) = \int d^{2} z_{\perp} G_{0}(x_{\perp} - z_{\perp}) G_{0}(y_{\perp} - z_{\perp})
\end{eqnarray}
and where we defined the quantity $\bar{\mu}^{2}(b^{+},a^{+}) \equiv \int_{a^{+}}^{b^{+}}dz^{+} \mu^{2}(z^{+})$. In the next subsections, we express the correlators appearing in the $\eta'$ cross section in terms of $\bar{U}$ and $\bar{V}$ which are defined in Eqs. (\ref{eq:one_av}) and (\ref{eq:two_av}).

\subsection{1 Wilson line - 1 color charge density correlator: the main building block}
\label{subsec:1W1cav}

The first correlator does not appear explicitly in the $\eta'$ production cross section but it is useful to understand the other results. We want to evaluate
\begin{eqnarray}
\label{eq:first_av}
F^{1,1}(b^{+},a^{+}) \equiv \langle U_{ab}(b^{+},a^{+}|x_{1\perp}) \rho_{c_{1}}(y_{1}^{+},y_{1\perp})  \rangle.
\end{eqnarray}
where we assume that $b^{+} > y_{1}^{+} > a^{+}$. For clarity, we define the sources included in Wilson lines as \textit{internal} sources as opposed to \textit{external} sources which appear explicitly in the correlator (like $\rho_{c_{1}}(y_{1}^{+},y_{1\perp})$ in Eq. (\ref{eq:first_av})).

The first step is to expand the Wilson line. The expression can then be written as
\begin{eqnarray}
F^{1,1}(b^{+},a^{+}) &=& \sum_{n=0}^{\infty} (-g^{2})^{n} \prod_{i=1}^{n} \int d^{2}z_{i \perp} G_{0}(x_{1\perp} - z_{i \perp}) (f_{a_{1}} ... f_{a_{n}})_{ab} \nonumber \\
&& \times \int_{a^{+}}^{b^{+}} dz_{1}^{+} \int_{a^{+}}^{z_{1}^{+}} dz_{2}^{+}... \int_{a^{+}}^{z_{n-1}^{+}} dz_{n}^{+} \nonumber \\
&& \times \langle \rho_{c_{1}}(y_{1}^{+},y_{1\perp})\rho_{a_{1}}(z_{1}^{+},z_{1 \perp}) ... \rho_{a_{n}}(z_{n}^{+},z_{n \perp}) \rangle 
\end{eqnarray}
where we used $t^{a}_{bc} = -if_{abc}$ and where $f_{abc}$ is the antisymmetric $SU(N_{c})$ structure constant. Using Wick theorem, the n-point correlation function can be expressed in terms of 2-point functions as the sum of all possible contractions. At first, we look only at the contractions of $\rho_{c_{1}}(y_{1}^{+},y_{1\perp})$  with  $\rho_{a_{1}}(z_{1}^{+},z_{1 \perp}) ... \rho_{a_{n}}(z_{n}^{+},z_{n \perp})$. Using the MV expression for the two point function Eq. (\ref{eq:MV}), we get
\begin{eqnarray}
F^{1,1}(b^{+},a^{+}) &=& \mu^{2}(y_{1}^{+})  G_{0}(x_{1\perp} - y_{1\perp})   \sum_{n=0}^{\infty} (-g^{2})^{n} \sum_{j=1}^{n} \left[ \prod_{i=1,i\neq j}^{n} \int d^{2}z_{i \perp} G_{0}(x_{1\perp} - z_{i \perp}) \right] \nonumber \\
&& \times (f_{a_{1}} ... f_{a_{j-1}}f_{c_{1}}f_{a_{j+1}}  ... f_{a_{n}})_{ab} \nonumber \\
&& \times \int_{a^{+}}^{b^{+}} dz_{1}^{+} \int_{a^{+}}^{z_{1}^{+}} dz_{2}^{+}... \int_{y_{1}^{+}}^{z_{j-2}^{+}} dz_{j-1}^{+} \int_{a^{+}}^{y_{1}^{+}} dz_{j+1}^{+}  ...  \int_{a^{+}}^{z_{n-1}^{+}} dz_{n}^{+} \nonumber \\
&& \times \langle \rho_{a_{1}}(z_{1}^{+},z_{1 \perp})... \rho_{a_{j-1}}(z_{j-1}^{+},z_{j-1 \perp})\rho_{a_{j+1}}(z_{j+1}^{+},z_{j+1 \perp}) ...  \rho_{a_{n}}(z_{n}^{+},z_{n \perp}) \rangle .
\end{eqnarray} 
\begin{figure}
\centering
		\includegraphics{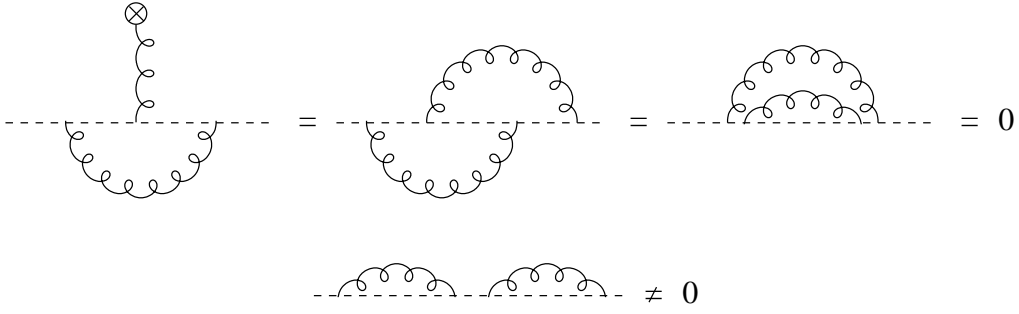}
	\caption{These are the different types of possible contractions using Wick theorem and path ordering. The first type is a contraction like $\langle \rho_{a_{j-1}}(z_{j-1}^{+},z_{j-1 \perp})\rho_{a_{j+1}}(z_{j+1}^{+},z_{j+1 \perp}) \rangle \langle \rho_{c}(y^{+},y_{\perp})\rho_{a_{j}}(z_{j}^{+},z_{j \perp}) \rangle  $ where $\rho_{c}$ is an external source. The other ones have only internal sources. The second one is like $\langle \rho_{a_{j}}(z_{j}^{+},z_{j \perp})\rho_{a_{j+2}}(z_{j+2}^{+},z_{j+2 \perp}) \rangle \langle \rho_{a_{j+1}}(z_{j+1}^{+},z_{j+1\perp})\rho_{a_{j+3}}(z_{j+3}^{+},z_{j+3 \perp}) \rangle  $, the third one is like $\langle \rho_{a_{j}}(z_{j}^{+},z_{j \perp})\rho_{a_{j+3}}(z_{j+3}^{+},z_{j+3 \perp}) \rangle \langle \rho_{a_{j+1}}(z_{j+1}^{+},z_{j+1\perp})\rho_{a_{j+2}}(z_{j+2}^{+},z_{j+2 \perp}) \rangle  $ and the last one is like $\langle \rho_{a_{j}}(z_{j}^{+},z_{j \perp})\rho_{a_{j+1}}(z_{j+1}^{+},z_{j+1 \perp}) \rangle \langle \rho_{a_{j+2}}(z_{j+2}^{+},z_{j+2\perp})\rho_{a_{j+3}}(z_{j+3}^{+},z_{j+3 \perp}) \rangle  $. Only the last one has a support and thus, a non-zero contribution.}
	\label{fig:zero_contribution}
\end{figure}
As argued in \cite{Gelis:2001da,Fukushima:2007dy}, only adjacent sources can be contracted due to the ordering in $z^{+}$. All the other nested and overlapping contractions have no support. It can also be shown that the contraction $\langle \rho_{a_{j-1}}(z_{j-1}^{+},z_{j-1 \perp})\rho_{a_{j+1}}(z_{j+1}^{+},z_{j+1 \perp}) \rangle  $ (where $\rho_{a_{j}}(z_{j}^{+},z_{j \perp})$ is contracted with the external source) does not have support either. These properties are shown diagrammatically in Fig. \ref{fig:zero_contribution}. They can be used to split the correlator in two parts like
\begin{eqnarray}
\langle \rho_{a_{1}}(z_{1}^{+},z_{1 \perp})... \rho_{a_{j-1}}(z_{j-1}^{+},z_{j-1 \perp})\rho_{a_{j+1}}(z_{j+1}^{+},z_{j+1 \perp}) 
 ... \rho_{a_{n}}(z_{n}^{+},z_{n \perp}) \rangle  = \nonumber \\
 \langle \rho_{a_{1}}(z_{1}^{+},z_{1 \perp})... \rho_{a_{j-1}}(z_{j-1}^{+},z_{j-1 \perp}) \rangle  \langle\rho_{a_{j+1}}(z_{j+1}^{+},z_{j+1 \perp})  ... \rho_{a_{n}}(z_{n}^{+},z_{n \perp}) \rangle .
\end{eqnarray}
By using these properties and by reorganizing the series, we have 
\begin{eqnarray}
F^{1,1}(b^{+},a^{+}) &=& \mu^{2}(y_{1}^{+})  G_{0}(x_{1\perp} - y_{1\perp})  f_{c_{1}d d'} \nonumber \\
&& \times  \biggl\{ \sum_{l=0}^{\infty} (-g^{2})^{l}  \left[ \prod_{i=1}^{l} \int d^{2}z_{i \perp} G_{0}(x_{1\perp} - z_{i \perp}) \right] (f_{a_{1}} ... f_{a_{l}})_{a d} \nonumber \\
&& \; \; \; \; \; \times  \int_{y_{1}^{+}}^{b^{+}} dz_{1}^{+} \int_{y_{1}^{+}}^{z_{1}^{+}} dz_{2}^{+}... \int_{y_{1}^{+}}^{z_{l-1}^{+}} dz_{l}^{+}  \langle \rho_{a_{1}}(z_{1}^{+},z_{1 \perp}) ... \rho_{a_{l}}(z_{l}^{+},z_{l \perp}) \rangle  \biggr\} \nonumber \\
&& \times  \biggl\{ \sum_{m=0}^{\infty} (-g^{2})^{m}  \left[ \prod_{j=1}^{m} \int d^{2}w_{j \perp} G_{0}(x_{1\perp} - w_{j \perp}) \right] (f_{b_{1}} ... f_{b_{m}})_{d' b} \nonumber \\
&& \; \; \; \; \; \times  \int_{a^{+}}^{y_{1}^{+}} dw_{1}^{+} \int_{a^{+}}^{w_{1}^{+}} dw_{2}^{+}... \int_{a^{+}}^{w_{m-1}^{+}} dw_{m}^{+}  \langle \rho_{b_{1}}(w_{1}^{+},w_{1 \perp}) ... \rho_{b_{m}}(w_{m}^{+},w_{m \perp}) \rangle  \biggr\}.
\end{eqnarray} 
This complicated expression is just a combination of Wilson lines that is given more succinctly as
\begin{eqnarray}
F^{1,1}(b^{+},a^{+}) &=& \mu^{2}(y_{1}^{+})  G_{0}(x_{1\perp} - y_{1\perp})  f_{c_{1} d d'} 
 \langle U_{ad}(b^{+},y_{1}^{+}|x_{1\perp}) \rangle \langle U_{d' b}(y_{1}^{+},a^{+}|x_{1\perp}) \rangle .
\end{eqnarray}
This can be simplified further by using Eq. (\ref{eq:one_av}) and the fact that $\bar{U}$ is an exponential. We can get easily that
\begin{eqnarray}
\label{eq:fin_av1}
F^{1,1}(b^{+},a^{+}) &=& \mu^{2}(y_{1}^{+})  G_{0}(x_{1\perp} - y_{1\perp})  f_{c_{1} ab} 
 \bar{U}(b^{+},a^{+}|x_{1\perp}) .
\end{eqnarray}
Specializing to the case of a charge distribution moving at the speed of light we have that $\rho_{a}(x^{+},x_{\perp}) = \delta(x^{+}) \rho_{a}(x_{\perp})$. Integrating both sides by $y^{+}$, we get
\begin{eqnarray}
\langle U_{ab}(b^{+},a^{+}|x_{1\perp}) \rho_{c_{1}} (y_{\perp})  \rangle
&=& \mu^{2}_{A}  G_{0}(x_{1\perp} - y_{1\perp})  f_{c_{1}ab} 
 \bar{U}(b^{+},a^{+}|x_{1\perp}). 
\end{eqnarray}

\subsection{1 Wilson line - 2 color charges densities correlator}

\begin{figure}
\centering
		\includegraphics[width=0.9\textwidth]{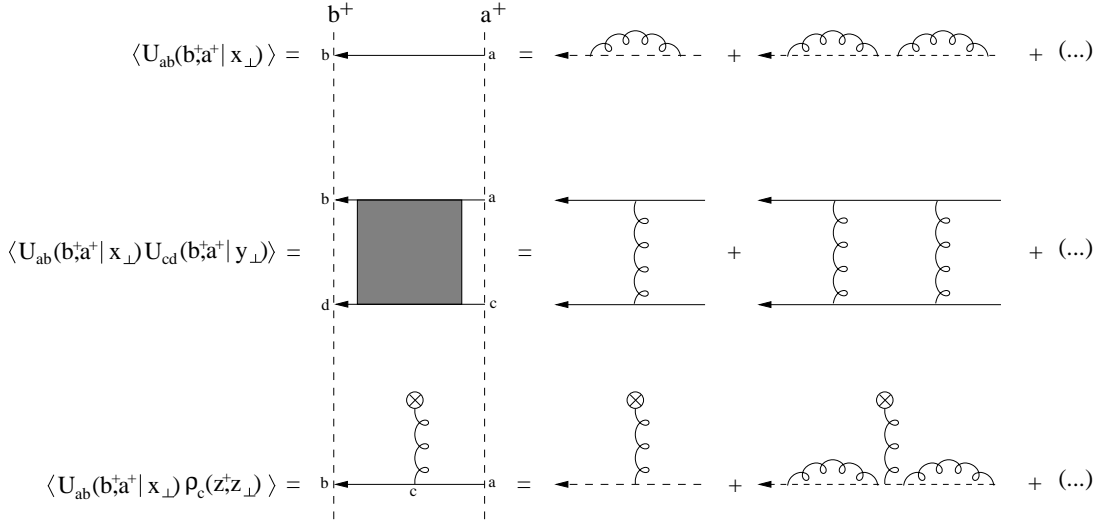}
	\caption{Diagrammatic rules shown with some of the first topologies they resum. Note that for the second correlator $\langle U_{ab}(b^{+},a^{+}|x_{\perp}) U_{cd}(b^{+},a^{+}|y_{\perp})  \rangle $, only the ladder-like diagram are allowed since all other topologies have no support and are zero \cite{Gelis:2001da,Fukushima:2007dy}. Finally, in this notation, the light-cone coordinates are ordered following the arrow, from left (the smallest) to right (the biggest) so that $b^{+}>a^{+}$.  }
	\label{fig:feyn_rules}
\end{figure}

\begin{figure}
\centering
		\includegraphics{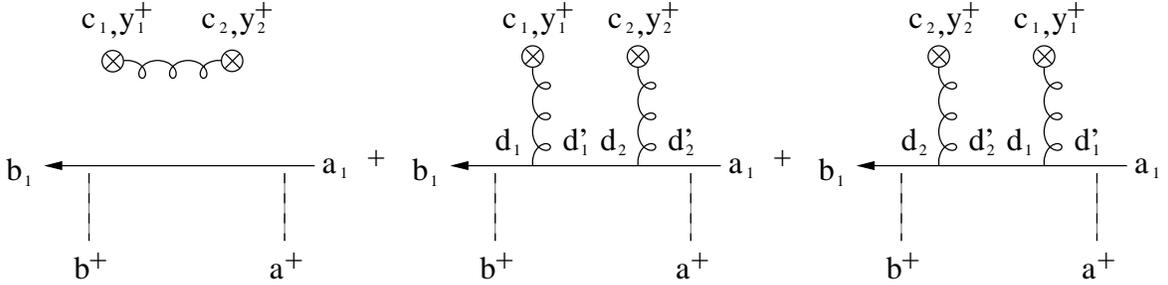}
	\caption{Diagrammatic representation of the 1 Wilson line - 2 color charge densities correlator $F^{2,1}(b^{+},a^{+})$. The first figure corresponds to the contraction of two \textit{external} sources. The other figures represent the cases where the external sources are contracted with \textit{internal} sources. The indices $a_{1},b_{1},c_{1},c_{2},d_{1},d'_{1},d_{2},d'_{2}$ are color indices while $a^{+},b^{+},y_{1}^{+},y_{2}^{+}$ are light-cone coordinates.}
	\label{fig:second_correlator}
\end{figure}

In this subsection, we compute the correlator
\begin{eqnarray}
F^{2,1}(b^{+},a^{+}) \equiv \langle U_{a_{1}b_{1}}(b^{+},a^{+}|x_{1\perp}) \rho_{c_{1}}(y_{1}^{+},y_{1\perp}) \rho_{c_{2}}(y_{2}^{+},y_{2\perp}) \rangle .
\end{eqnarray}
It is possible to devise diagrammatic rules shown in Fig. \ref{fig:feyn_rules} that can be used to write $F^{2,1}(b^{+},a^{+})$ in terms of known quantities. The rules are discussed in more details and generalized to all cases in App. \ref{app:dia_rules}. Using these results, $F^{2,1}(b^{+},a^{+})$ can be represented diagrammatically as in Fig. \ref{fig:second_correlator}. This includes all possible contractions and topologies that need to be resummed. According to the rules of App. \ref{app:dia_rules}, it can be written as
\begin{eqnarray}
F^{2,1}(b^{+},a^{+}) &=& \langle \rho_{c_{1}}(y_{1}^{+},y_{1\perp}) \rho_{c_{2}}(y_{2}^{+},y_{2\perp}) \rangle \langle U_{a_{1}b_{1}}(b^{+},a^{+}|x_{1\perp}) \rangle \nonumber \\
&+& \theta(y_{1}^{+}-y_{2}^{+}) \mu^{2}(y_{1}^{+}) \mu^{2}(y_{2}^{+}) f_{c_{1}d_{1}d'_{1}} f_{c_{2}d_{2}d'_{2}} G_{0}(y_{1\perp} - x_{1 \perp}) G_{0}(y_{2\perp} - x_{1 \perp}) \\ \nonumber 
&& \times \langle U_{b_{1}d_{1}}(b^{+},y_{1}^{+}|x_{1\perp}) \rangle \langle U_{d'_{1}d_{2}}(y_{1}^{+},y_{2}^{+}|x_{1\perp}) \rangle \langle U_{d'_{2}a_{1}}(y_{2}^{+},a^{+}|x_{1\perp}) \rangle\nonumber \\
&+& \theta(y_{2}^{+}-y_{1}^{+}) \mu^{2}(y_{1}^{+}) \mu^{2}(y_{2}^{+}) f_{c_{1}d_{1}d'_{1}} f_{c_{2}d_{2}d'_{2}} G_{0}(y_{1\perp} - x_{1 \perp}) G_{0}(y_{2\perp} - x_{1 \perp}) \\ \nonumber 
&& \times \langle U_{b_{1}d_{2}}(b^{+},y_{2}^{+}|x_{1\perp}) \rangle \langle U_{d'_{2}d_{1}}(y_{2}^{+},y_{1}^{+}|x_{1\perp}) \rangle \langle U_{d'_{1}a_{1}}(y_{1}^{+},a^{+}|x_{1\perp}) \rangle.
\end{eqnarray}
This can be simplified further by using the explicit expressions given in Eqs. (\ref{eq:MV}), (\ref{eq:one_av}) and (\ref{eq:fin_av1}). We get that
\begin{eqnarray}
F^{2,1}(b^{+},a^{+}) &=& \delta_{c_{1}c_{2}} \delta_{a_{1}b_{1}} \mu^{2}(y_{1}^{+})\delta(y_{1}^{+} - y_{2}^{+}) \delta^{2}(y_{1\perp} - y_{2\perp})  \bar{U}(b^{+},a^{+}|x_{1\perp})  \nonumber \\
&+& \mu^{2}(y_{1}^{+}) \mu^{2}(y_{2}^{+})  G_{0}(x_{1\perp} - y_{1\perp}) G_{0}(x_{1\perp} - y_{2\perp})  \bar{U}(b^{+},a^{+}|x_{1\perp}) \nonumber \\
&& \times (f_{c_{1} b_{1} d} f_{c_{2} d a_{1}}\theta(y_{1}^{+} - y_{2}^{+}) + f_{c_{2} b_{1} d} f_{c_{1} d a_{1}}\theta(y_{2}^{+} - y_{1}^{+})) 
\end{eqnarray}
Considering that the nuclei is moving at the speed of light and integrating on both sides by $y_{1}^{+}$ and $y_{2}^{+}$, we get
\begin{eqnarray}
\langle U_{ab}(b^{+},a^{+}|x_{1\perp}) \rho_{c_{1}}(y_{1\perp}) \rho_{c_{2}}(y_{2\perp}) \rangle
&=& \delta_{c_{1}c_{2}} \delta_{a_{1}b_{1}} \mu^{2}_{A} \delta^{2}(y_{1\perp} - y_{2\perp})  \bar{U}(b^{+},a^{+}|x_{1\perp})  \nonumber \\
&+& \mu_{c}^{4}  G_{0}(x_{1\perp} - y_{1\perp}) G_{0}(x_{1\perp} - y_{2\perp})  \bar{U}(b^{+},a^{+}|x_{1\perp}) \nonumber \\
&& \times (f_{c_{1} b_{1} d} f_{c_{2} d a_{1}} + f_{c_{2} b_{1} d} f_{c_{1} d a_{1}})  
\end{eqnarray}
where we defined
\begin{eqnarray}
\mu_{c}^{4} = \int_{-\infty}^{\infty} du^{+}\int_{u^{+}}^{\infty} dv^{+} \mu^{2}(u^{+}) \mu^{2}(v^{+}).
\end{eqnarray}

\subsection{2 Wilson lines - 2 color charges densities correlator}

\begin{figure}
\centering
		\includegraphics[width=0.7\textwidth]{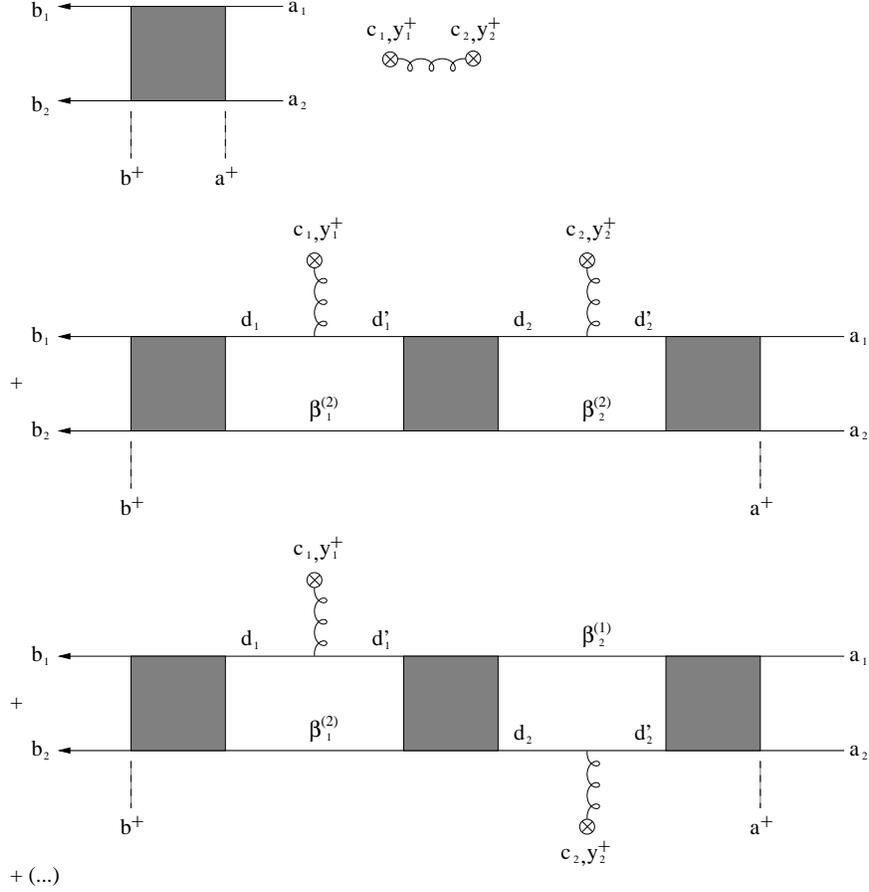}
	\caption{These are the first few diagrams included in the calculation of $F^{2,2}(b^{+},a^{+})$. They differ essentially by the way the sources are inserted between the blobs. }
	\label{fig:correlatorUUpp}
\end{figure}

In this subsection, we compute the correlator
\begin{eqnarray}
F^{2,2}(b^{+},a^{+}) \equiv \langle U_{a_{1} b_{1}}(b^{+},a^{+}|x_{1 \perp}) U_{a_{2} b_{2}}(b^{+},a^{+}|x_{2 \perp}) \rho_{c_{1}}(y_{1}^{+},y_{1\perp}) \rho_{c_{2}}(y_{2}^{+},y_{2\perp}) \rangle .
\end{eqnarray}
This calculation is similar to the one of the second correlator $F^{2,1}$ using the diagrammatic rules. Three typical diagrams of $F^{2,2}(b^{+},a^{+})$ out of seven are shown in Fig. \ref{fig:correlatorUUpp}. The first one corresponds to the contraction of the two \textit{external} sources and the other ones corresponds the connected part which consists in all contractions between \textit{internal} and \textit{external} sources. 

The first diagram is straightforward to compute. It is given by
\begin{eqnarray}
F^{2,2}_{1}(b^{+},a^{+}) \equiv \langle  U_{a_{1} b_{1}}(b^{+},a^{+}|x_{1 \perp}) U_{a_{2} b_{2}}(b^{+},a^{+}|x_{2 \perp})\rangle \langle \rho_{c_{1}}(y_{1}^{+},y_{1\perp}) \rho_{c_{2}}(y_{2}^{+},y_{2\perp}) \rangle .
\end{eqnarray}
Using Eqs. (\ref{eq:MV}) and (\ref{eq:two_av}), we get that 
\begin{eqnarray}
F^{2,2}_{1}(b^{+},a^{+}) &=& \mu^{2}(y_{1}^{+}) \delta(y_{1}^{+} - y_{2}^{+}) \delta^{2}(y_{1\perp} - y_{2\perp}) \bar{V} (b^{+},a^{+}|x_{1 \perp},x_{2 \perp}) \frac{ \delta_{c_{1}c_{2}} \delta_{a_{1} a_{2}} \delta_{b_{1} b_{2}}}{N_{c}^{2}-1} .
\end{eqnarray}

The second diagram shown in Fig. \ref{fig:correlatorUUpp} is given by
\begin{eqnarray}
F^{2,2}_{2}(b^{+},a^{+}) &=& \theta(y_{1}^{+}-y_{2}^{+}) \mu^{2}(y_{1}^{+}) \mu^{2}(y_{2}^{+}) f_{c_{1}d_{1}d'_{1}} f_{c_{2}d_{2}d'_{2}} G_{0}(y_{1\perp} - x_{1 \perp}) G_{0}(y_{2\perp} - x_{1 \perp}) \\ \nonumber 
&& \times \langle U_{b_{1}d_{1}}(b^{+},y_{1}^{+}|x_{1\perp}) U_{b_{2}\beta_{1}^{(2)}}(b^{+},y_{1}^{+}|x_{2\perp}) \rangle \langle U_{d'_{1}d_{2}}(y_{1}^{+},y_{2}^{+}|x_{1\perp}) U_{\beta_{1}^{(2)}\beta_{2}^{(2)}}(y_{1}^{+},y_{2}^{+}|x_{2\perp})\rangle \nonumber \\
&& \times  \langle U_{d'_{2}a_{1}}(y_{2}^{+},a^{+}|x_{1\perp}) U_{\beta_{2}^{(2)}a_{2}}(y_{2}^{+},a^{+}|x_{2\perp})\rangle.
\end{eqnarray}
Using Eqs. (\ref{eq:one_av}) and (\ref{eq:two_av}), we get
\begin{eqnarray}
F^{2,2}_{2}(b^{+},a^{+}) &=& \theta(y_{1}^{+}-y_{2}^{+}) \mu^{2}(y_{1}^{+}) \mu^{2}(y_{2}^{+}) f_{c_{1}d_{1}d_{1}} f_{c_{2}d_{2}d_{2}} G_{0}(y_{1\perp} - x_{1 \perp}) G_{0}(y_{2\perp} - x_{1 \perp}) \\ \nonumber 
&& \times \bar{V}(b^{+},y_{1}^{+}|x_{1\perp},x_{2\perp}) \bar{V}(y_{1}^{+},y_{2}^{+}|x_{1\perp},x_{2\perp}) \bar{V}(y_{2}^{+},a^{+}|x_{1\perp},x_{2\perp}) \nonumber \\
&=& 0
\end{eqnarray}
which is zero because of the color structure. All the other diagrams included in $F^{2,2}(b^{+},a^{+})$ can be computed in a similar way. There are five other different ways of inserting the source and they all vanish because of  the color structure. The only non-zero term is the first one and so we have $F^{2,2}(b^{+},a^{+})=F^{2,2}_{1}(b^{+},a^{+})$. Finally, with a nucleus moving at the speed of light, we obtain
\begin{eqnarray}
\label{eq:ave_3_final}
\langle U_{a_{1} b_{1}}(b^{+},a^{+}|x_{1 \perp}) U_{a_{2} b_{2}}(b^{+},a^{+}|x_{2 \perp}) \rho_{c_{1}}(y_{1\perp}) \rho_{c_{2}}(y_{2\perp}) \rangle
&=&  \mu_{A}^{2}  \delta^{2}(y_{1\perp} - y_{2\perp})  \frac{\delta_{c_{1}c_{2}} \delta_{a_{1} a_{2}} \delta_{b_{1} b_{2}}}{N_{c}^{2}-1} \nonumber \\
&& \times \bar{V} (b^{+},a^{+}|x_{1 \perp},x_{2 \perp}) .
\end{eqnarray}
This concludes the computation of correlators. We are now in a position to evaluate the cross section within the MV model.

\section{Numerical evaluation of the cross section}
\label{sec:num_eval}

\subsection{Proton-proton case}

In this section, we evaluate the cross section numerically in $pp$ collisions. This is done by using the expression of the cross section given by Eq. (\ref{eq:cross_kfac}) which can be obtained either from the dilute limit of the $pA$ result or from $k_{\perp}$-factorization techniques (see Appendix \ref{app:pp_k_perp}). In $pp$ collisions, the cross section is related to uPDF that describe the distribution of gluons inside each protons. There exist many parametrizations of these distribution functions differing mainly in the way the evolution equation is solved. Among the most successful ones are (the description of these parametrizations can be found in \cite{Andersson:2002cf} \footnote{We would like to thanks H. Jung for
handing us his FORTRAN routine CAUNIGLU which evaluates numerically all of
these parametrizations. It can be found at
http://www.desy.de/~jung/cascade/updf.html.} ):
\begin{itemize}

\item DIG (Derivative of the Integrated Gluon distribution
function)

\item CCFM (Catani, Ciafaloni, Fiorani, Marchesini) \cite{Ciafaloni:1987ur,Catani:1989sg,Catani:1989yc,Jung:2000hk,Jung:2001rp,Jung:2003wu}

\item KMR (Kimber, Martin, Ryskin) \cite{Kimber:2001sc}
\end{itemize} 
These parametrizations are used to compute the cross section for $\eta'$ production in $pp$ collisions and to compare with the result for $pA$ collisions at small saturation scale.

\begin{figure}
\centering
		\includegraphics[width=0.7\textwidth]{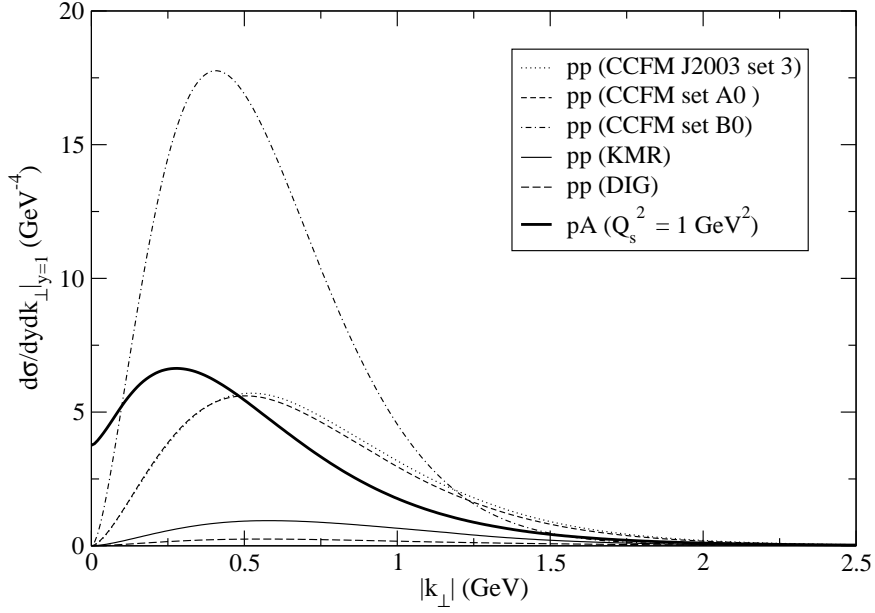}
	\caption{Numerical results of the inclusive differential cross section at rapidity ($y=1$) and at RHIC energy ($\sqrt{s}$=200 GeV). The results for $pp$ collisions are scaled by the number of nucleons $A$ to make a comparison with $pA$. The cross section for $pA$ collisions is evaluated at the saturation scale $Q_{s} = 1 \; \mbox{GeV}$.  }
	\label{fig:results_pp_y1}
\end{figure}

The final result is obtained by integrating Eq. (\ref{eq:cross_kfac}) using the VEGAS and the CUHRE algorithms implemented in the CUBA package \cite{Hahn:2004fe}. The number of color is set to $N_{c}=3$, the
center of mass energy to $\sqrt{s} \approx 200 \; \mbox{GeV}$ (RHIC) and
the mass of $\eta'$ to $M=0.957 \; \mbox{GeV}$. To make a comparison with $pA$ collisions for a wide range of transverse momentum where the MV model is valid (see Fig. (\ref{fig:kin_range})), we chose the rapidity $y=1$. The results of the numerical calculation are shown in Fig. (\ref{fig:results_pp_y1}) where we also present the result for the $pA$ case at small saturation scale ($Q_{s}^{2} = 1 \; \mbox{GeV}$) and with the proton described by the parametrization CCFM J2003 set 3.

\subsection{Proton-nucleus case using the MV model}

We now consider $\eta'$ production for $pA$ collisions for which the cross section is given by Eq. (\ref{eq:cross01}). This formal expression can be simplified by using the results of section \ref{sec:averages} where the correlators of Wilson lines are evaluated in the MV model. Note here that in the MV model, the nucleus is considered as an infinite source of charge in the transverse plane so there are no edge effects. In this kind of description, translation invariance in the transverse plane is preserved and therefore the correlators have the following property
\begin{eqnarray}
\label{eq:trans}
\langle \rho_{A,a} (x_{\perp}) \rho_{A,b} (y_{\perp})U_{ce} (z_{\perp} ) U_{c'e'} (w_{\perp})  \rangle = \langle \rho_{A,a} (x_{\perp}-w_{\perp}) \rho_{A,b} (y_{\perp}-w_{\perp})U_{ce} (z_{\perp}-w_{\perp} ) U_{c'e'} (0)  \rangle.
\end{eqnarray} 
Then, we can write
\begin{eqnarray}
(2\pi)^3 2E_k \frac{d\sigma}{ d^3 k}
&\approx&
\frac{g^{2} \pi^{2}}{(N_{c}^{2}-1)}  \int \frac{d^{2}p_{\perp} d^{2}q_{\perp} d^{2}r_{\perp} }{(2\pi)^{6}} F(-p_{\perp}^{2} , -p_{2,\perp}^{2}) F^{*}(-q_{\perp}^{2},-q_{2,\perp}^{2}) \nonumber \\
&& \times \biggl[ \langle \rho_{A,a}^{*}(p_{\perp}) \rho_{A,a}(q_{\perp}) \rangle (2\pi)^{2} \delta^{2}(p_{2,\perp} - r_{\perp})  (2\pi)^{2} \delta^{2}( q_{2,\perp} - r_{\perp}) \nonumber \\
&&+  \int d^{2}x_{\perp} d^{2}y_{\perp} d^{2}z_{\perp} d^{2}w_{\perp}   e^{ip_{\perp} \cdot x_{\perp} - iq_{\perp} \cdot y_{\perp} +i( p_{2,\perp} - r_{\perp}) \cdot z_{\perp} - i( q_{2,\perp} - r_{\perp})\cdot w_{\perp}}\nonumber \\
&& \times  \langle \rho_{A,a} (x_{\perp}) \rho_{A,b} (y_{\perp})U_{ac} (z_{\perp} ) U_{bc} (w_{\perp})  \rangle \biggr] \epsilon_{ij}\epsilon_{kl} \frac{p^{i} q^{k} r^{j} r^{l}}{p^{2}_{\perp} q^{2}_{\perp} r_{\perp}^{2} }   \phi_{p}(r_{\perp}^{2},x).
\end{eqnarray}
To obtain this expression, we used Eq. (\ref{eq:coor_unint_p}) to convert the average on proton sources to an unintegrated distribution function and we Fourier transformed the correlator in the second term. We also neglected all terms with only one Wilson line because they are numerically very small compared to the other terms. This can be seen as follows. First, using translation invariance of the correlator, we see that averages with one Wilson line in the cross section are proportional to $\bar{U}(0)$ (see Eq. (\ref{eq:trans})). This quantity is small because there is an infrared singularity appearing in the argument of the exponential in the following way \cite{Fukushima:2007dy}:
\begin{eqnarray}
\label{eq:U_approx}
 \bar{U}(0)  
&=&  \exp \left[ -\frac{N_{c}}{2} L(0,0) Q_{s}  \right]=\exp \left[ -\frac{N_{c}}{2} Q_{s}^{2} \int \frac{d^{2}p_{\perp}}{(2\pi)^{2}} \frac{1}{p_{\perp}^{4}} \right] \approx 0.
\end{eqnarray}
This is not exactly zero because the infrared singularity is regulated by non-perturbative effects (confinement) and these induce a cutoff of order $\Lambda_{\rm{QCD}}$. The saturation scale in a large nuclei at small-$x$ satisfies $Q_{s}^{2} \gg \Lambda_{\rm{QCD}}^{2}$ so that  $\bar{U}(0) \sim \exp\left[-\frac{Q_{s}^{2}}{\Lambda_{\rm{QCD}}^{2}} \right] \ll 1$. Thus, the correlators with one Wilson line can be neglected.

It is then a straightforward calculation, involving some change of variables, translation invariance and Eqs. (\ref{eq:MV}) and (\ref{eq:ave_3_final}), to obtain
\begin{eqnarray}
\label{eq:simp_cross}
(2\pi)^3 2E_k \frac{d\sigma}{ d^3 k}
&=&
\pi^{2}g^{2} \mu^{2}_{A} S_{\perp}  \int \frac{d^{2}p_{\perp} d^{2}r_{\perp} }{(2\pi)^{4}} |F(-p_{\perp}^{2} , -p_{2,\perp}^{2})|^{2} \epsilon_{ij}\epsilon_{kl} \frac{p^{i} p^{k} r^{j} r^{l}}{p^{2}_{\perp} (p^{2}_{\perp}+ \Lambda^{2}) r_{\perp}^{2} }     \nonumber \\
&& \times  \phi_{p}(r_{\perp}^{2},x)  \biggl[ (2\pi)^{2} \delta^{2}( p_{2,\perp} - r_{\perp})  +    \int  d^{2}z_{\perp}     e^{i( p_{2,\perp} - r_{\perp}) \cdot z_{\perp}} \bar{V}(z_{\perp},0) \biggr]
\end{eqnarray}  
where $S_{\perp} = (2\pi)^{2}\delta^{2}(0)$ is interpreted as the transverse area of the nuclei. We also introduced here an infrared regulator given by $\Lambda$ in one of the denominator $\frac{1}{p_{\perp}^{2}}$. This is because gluons separated by a distance greater than the size of a nucleon ($\sim 1 \; \mbox{fm}$) are not correlated because of confinement so that $\Lambda = \Lambda_{\rm{QCD}}$. This is not taken into account explicitly in the MV model, leading to infrared divergent quantities \cite{Lam:1999wu} which are regulated by adding the infrared regulator \cite{Gyulassy:1997vt}. This is equivalent to define a gluon distribution function for the MV model as
\begin{eqnarray}
\phi_{A}(x,p_{\perp},Q^2) = \frac{\alpha_{s}(N_{c}^{2}-1)A}{2\pi} \frac{1}{p_{\perp}^{2}+\Lambda^{2}}.
\end{eqnarray}
We do not introduce a regulator in the other denominator $\frac{1}{p_{\perp}^{2}}$ because as seen in Eq. (\ref{eq:k_fact_cross}), it is part of the matrix element $|\mathcal{M}|^{2}  \sim \epsilon_{ij}\epsilon_{kl} \frac{p^{i} p^{k} r^{j} r^{l}}{p^{2}_{\perp}  r_{\perp}^{2} }$ which is well behaved in the infrared.

In the first term of the cross section, one integration can be done easily with the delta function. The remaining integrals will be done numerically. The second term can be simplified further by doing some integrals analytically as shown in Appendix \ref{app:simp}. The first term is very similar to the $pp$ cross section and does not involve any saturation scale dependence because it does not include any multiscatterings. The $Q_{s}$ dependence occurs in the second term through the expression of $\bar{V}$ where we define it as
\begin{eqnarray}
Q_{s}^{2} = \frac{N_{c} \mu_{A}^{2}g^{4}}{2\pi}.
\end{eqnarray}

\subsubsection{Kinematical range}

\begin{figure}
\centering
		\includegraphics[width=0.6\textwidth,angle=-90]{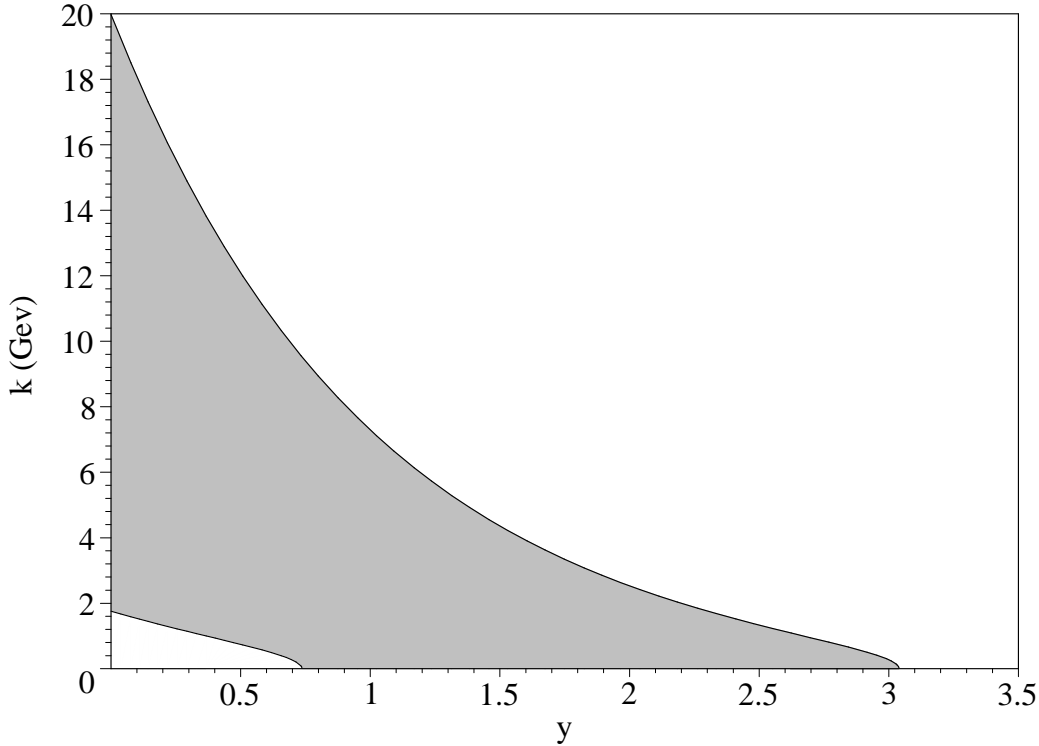}
	\caption{The grey region is the kinematic range where the calculation using the MV model is valid for RHIC energies in terms of the $\eta'$ transverse momentum $k=|k_{\perp}|$ and rapidity $y$. }
	\label{fig:kin_range}
\end{figure}

The cross section derived in the previous section is restricted to a certain range of validity because we are using the MV model. This model can be used when the radiative corrections which goes like $\alpha_{s}\ln(1/x)$ are not too large, which is when $0.01 \lesssim x_{A} \lesssim 0.1$. For $\eta'$ production, the momentum fraction of gluons in the proton and the nucleus are given by 
\begin{eqnarray}
x_{p}&=& \sqrt{\frac{M^{2}+k_{\perp}^{2}}{s}}e^{y}, \\
x_{A}&=& \sqrt{\frac{M^{2}+k_{\perp}^{2}}{s}}e^{-y}.
\end{eqnarray}
The kinematic range in terms of the $\eta'$ transverse momentum and rapidity where $0.01 \lesssim x_{A} \lesssim 0.1$ is depicted in Fig. \ref{fig:kin_range} for RHIC energies. 

It is possible to extend the range of validity to smaller values of $x_{A}$ by using the JIMWLK equation \cite{Iancu:2000hn,Ferreiro:2001qy,JalilianMarian:1996xn,JalilianMarian:1997gr,JalilianMarian:1997jx}. This renormalization group equation resums the large radiative corrections due to very small $x$ modes. The effect is to change the correlation between sources and in general, one looses the Gaussian structure of the weight functional $W[\rho]$. This complicates the computation of Wilson line correlators and is outside the scope of this article.

\subsubsection{Numerical Results}
\label{sec:num_res}

\begin{figure}[t]
\centering
		\includegraphics[width=0.7\textwidth]{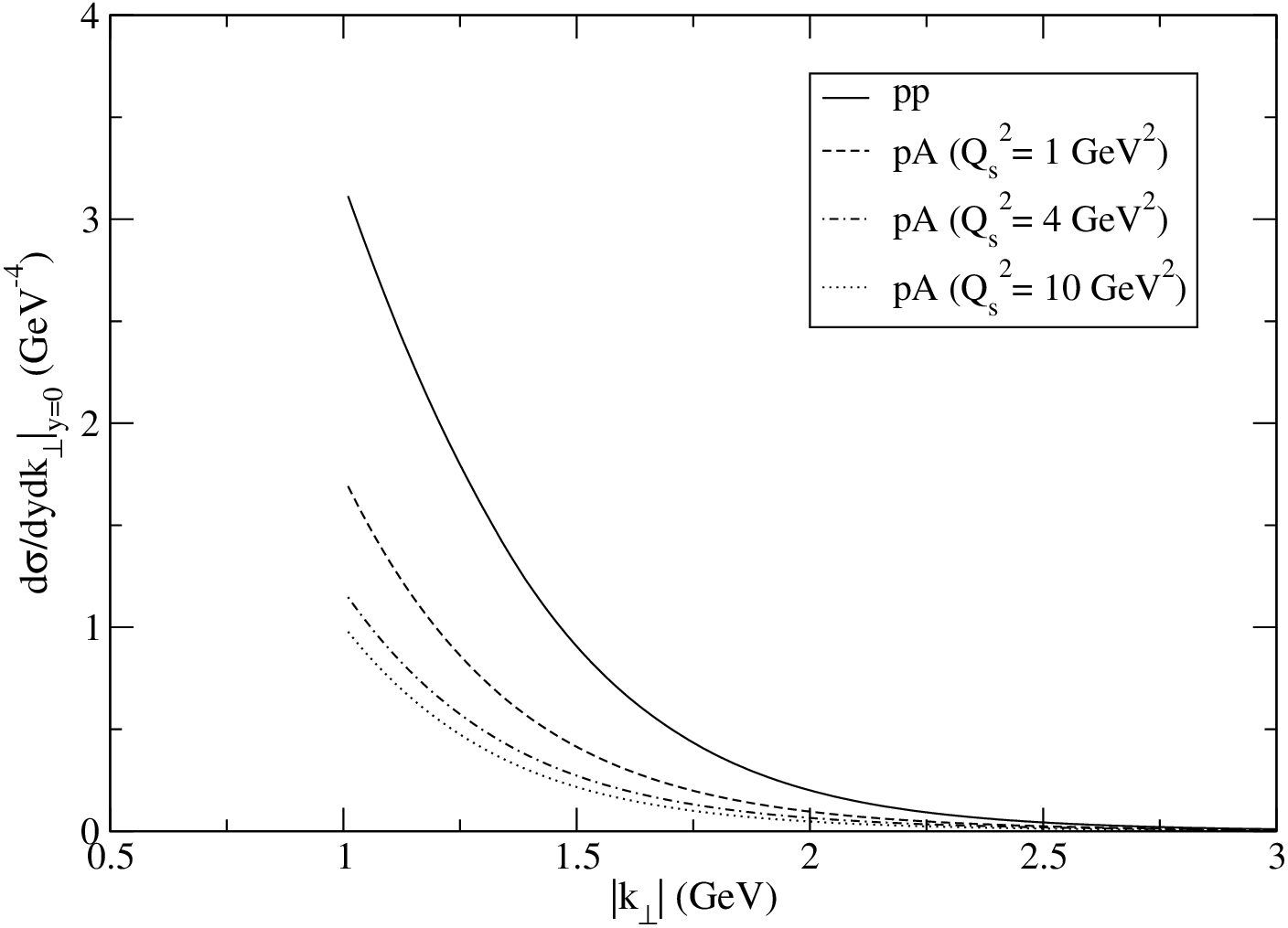}
	\caption{Numerical results of the inclusive differential cross section at midrapidity ($y=0$) and at RHIC energy ($\sqrt{s}$=200 GeV). The first curve is the result of the cross section for $pp$ collisions scaled by the number of nucleons $A$. The other curves are the results for the cross section for $pA$ collisions for different values of the saturation scale ($Q_{s}$=1,4 and 10 GeV).  }
	\label{fig:results_y0}
\end{figure}

\begin{figure}[t]
\centering
		\includegraphics[width=0.7\textwidth]{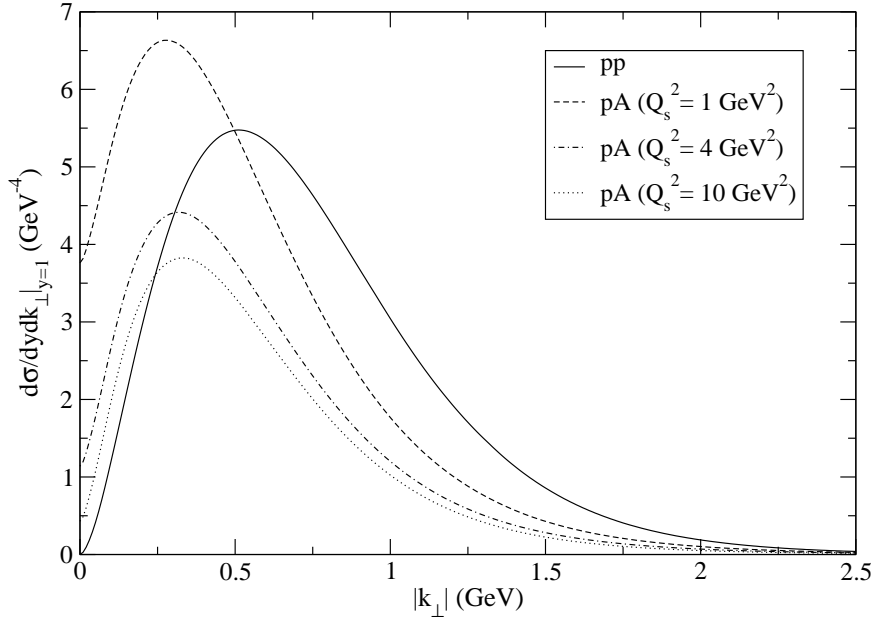}
	\caption{Numerical results of the inclusive differential cross section at rapidity $y=1$ and at RHIC energy ($\sqrt{s}$=200 GeV). The first curve is the result of the cross section for $pp$ collisions scaled by the number of nucleons $A$. The other curves are the results for the cross section for $pA$ collisions for different values of the saturation scale ($Q_{s}$=1,4 and 10 GeV).  }
	\label{fig:results_y1}
\end{figure}

We present in this section the numerical results for $\eta'$ production. We integrate numerically Eq. (\ref{eq:final_simp}) and the first term of Eq. (\ref{eq:simp_cross}) using the same values for the parameters as in $pp$ collisions. The strong coupling constant appearing in the cross section is evaluated at the $\eta'$ transverse mass scale $M_{\perp}^{2}$. The unintegrated distribution functions chosen for the protons is the CCFM J2003 set 3 because it is very successful in the description of other observables like charm and bottom production at the Tevatron \cite{Jung:2003wu}. The results of the numerical integration are shown in Figs. (\ref{fig:results_y0}) and (\ref{fig:results_y1}) for different values for the rapidity. 

\begin{figure}[t]
\centering
		\includegraphics[width=0.7\textwidth]{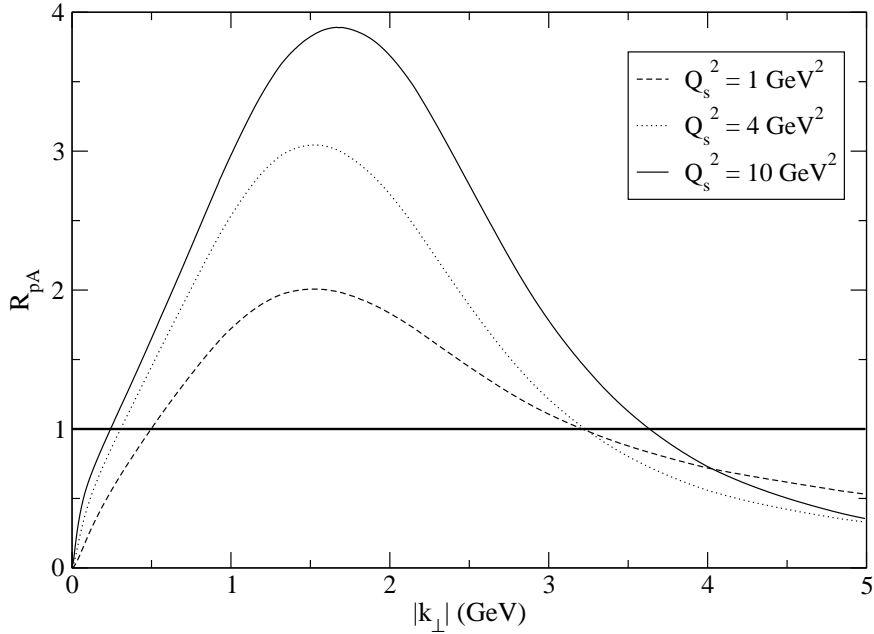}
	\caption{Numerical results for the inverse nuclear modification factor at rapidity ($y=1$) and at RHIC energy ($\sqrt{s}$=200 GeV). We show the results for different values of the saturation scale ($Q_{s}$=1,4 and 10 GeV).  }
	\label{fig:results_RpA_y1}
\end{figure}

In Fig. (\ref{fig:results_RpA_y1}), we show the result for the inverse nuclear modification factor $R_{pA}$ for $y=1$ defined as 
\begin{eqnarray}
\frac{1}{R_{pA}} = \frac{A\frac{d\sigma^{pp}}{d^{2}k_{\perp}dy}}{\frac{d\sigma^{pA}}{d^{2}k_{\perp}dy}}.
\end{eqnarray}
We use this because the $pp$ cross section is zero at $|k_{\perp}|=0$.

\subsection{Analysis}

For the $pp$ cross section, there is a wide range of variability for the cross section depending on the parametrization of the uPDF used. This feature can be used to discriminate between the different parametrizations to determine the most accurate one. Thus, by combining this analysis with experimental data, the $\eta'$ production becomes another observable that can be utilized to constrain models of uPDF. A similar conclusion was reached in \cite{FillionGourdeau:2007ee} where the $f_{2}$-meson production was studied. Note however that the results of the cross section at small transverse momentum should be treated carefully. As stated in the appendix \ref{app:pp_k_perp}, the validity of the $k_{\perp}$-factorization approach depends on the presence of a large scale compared to the QCD scale given in our case by $Q^{2} \sim M_{\perp}^{2}$. At very small momentum, we have that $\frac{M_{\perp}^{2}}{\Lambda_{\rm{QCD}}^2} \approx 23$ for which the semihard inequality $\Lambda_{\rm{QCD}}^{2} \ll Q^{2} \ll \sqrt{s}$ that guarantees the accuracy of $k_{\perp}$-factorization is only marginally satisfied. Thus, at small momentum, the cross section should be seen as an extrapolation of the $k_{\perp}$-factorized cross section to a regime where $k_{\perp}$-factorization cannot be rigorously proven.  

The $pA$ cross section shows a strong dependence on the saturation scale. As $Q_{s}$ becomes smaller, the magnitude of the cross section looks more like the $pp$ cross section. On the other hand, the cross section diminishes as $Q_{s}$ becomes larger. This can be understood in the following way. The saturation scale is defined as the momentum at which the probability of interaction between different parton cascades is of order one \cite{Iancu:2002xk,Iancu:2003xm}. The partons having a transverse size satisfying $\delta x_{\perp} \sim \frac{1}{Q_{\perp}} \geq \frac{1}{Q_{s}}$ will have a very high probability to recombine. As the saturation scale is increased, there will be more gluons that will be sensitive to these nonlinear effects. The final result is that as more and more gluons have a high probability to recombine, the growth of their population will decrease and so is the cross section for a given set of parameters. This effect is clearly seen in the cross section computed in our model. The fact that the $\eta'$ production cross section is sensitive to the saturation scale can be used to estimate the numerical value of $Q_{s}$. The current estimate based on HERA data for deep inelastic scattering shows that for RHIC energy, it is given by $Q_{s} \sim 1 - 2 \;\mbox{GeV}$ \cite{Iancu:2002xk,Iancu:2003xm}. It would be interesting to compare our calculation to experimental data to make an estimation of $Q_{s}$ based on $\eta'$ production and see if it is consistent with the previous estimation. 

The nuclear modification factor shows clearly the saturation effects for $0.4 \lesssim |k_{\perp}| \lesssim 3.4$ because in that range, $R_{pA}>1$. At large transverse momentum, it approaches the value $R_{pA} \approx 0.45$. One would expect $R_{pA}$ to be one in that range of momentum because it corresponds to the regime where there are no saturation effects and where the $k_{\perp}$-factorized cross section can be used. This discrepancy can be explained by looking at the approximations made during the calculation. We neglect two terms in the cross section that are proportional to one Wilson line (see Eq. (\ref{eq:U_approx})). In the dilute limit (when $|k_{\perp}| > Q_{s}$), these terms cannot be neglected and have a non zero contribution that would make $R_{pA} \approx 1$. In that sense, our approximation fails at large transverse momentum and our result should be treated carefully in that regime.

\section{Conclusion}

In this article, the inclusive cross section for $\eta'$ production in $pp$ and $pA$ collisions is computed. The $pp$ case is analyzed for different reasons. First, it serves as a basis to validate our model for $\eta'$ production against experimental data. This model includes two main ingredients. The first one is the effective theory which is at the foundation of our analysis. We use in our study a very simplified form of this effective theory where the vertex is given by Eq. (\ref{eq:vertex2}). As shown in \cite{Atwood:1997bn,Ali:2003kg,Ali:2000ci,Muta:1999tc,Ahmady:1998mi,Agaev:2002ek,Kroll:2002nt}, this is a very crude approximation of the real vertex, so there are some improvements that can be done in this direction in the future. The second ingredient are the uPDF parametrizations. As seen in Fig. (\ref{fig:results_pp_y1}), there is still a large variability in the predictions made by different uPDF. Therefore, $\eta'$ production could be used to constrain the models of uPDF once it is compared to experimental data. The same conclusion was obtained in \cite{Szczurek:2006bn} for the exclusive process $p+p \rightarrow p+p+\eta'$. Finally, from the theoretical point of view, we studied $pp$ collisions to see if the cross section can be obtained as the low density limit of the $pA$ cross section. There are now many known examples where this can be seen like gluon production \cite{Gyulassy:1997vt,Kovchegov:1997ke,Blaizot:2004wu}, quark production \cite{Blaizot:2004wv,Gelis:2003vh} and tensor meson production \cite{FillionGourdeau:2007ee}. We have shown that $\eta'$ production also obeys this property and in that sense, it is a consistency check for the approach used for the $pA$ case.

The $pA$ results for the $\eta'$ inclusive cross section show some very interesting features. First, we show that they are sensitive to the value of the saturation scale. This can also be seen in the plot of the nuclear modification factor. This property could be used to make an estimate of $Q_{s}$ by comparing with experimental data, which is one of the main goal of our analysis. This information is very important for the study of other particle production such as quark and gluon production in both $pA$ and $AA$ where saturation effects play an important role. Thus, a measurement of $\eta'$ at RHIC would improve our knowledge of gluon distribution in a nucleus.

Throughout the article, we assumed that the gluon fusion process was the dominant mechanism in $\eta'$ production. There is one other production process that could also be important. It is the photon fusion where two off-shell photons emitted by the protons or the nucleus merge to give a $\eta'$ such as $\gamma^{*} + \gamma^{*} \rightarrow \eta' +X$. This is estimated in \cite{Szczurek:2006bn} and according to this analysis, it should be subdominant in the exclusive production. We assume that this holds also in our study, although a careful analysis of this process should be performed. Of course, our calculation could be made more accurate by investigating this last issue.

\begin{acknowledgments}

The authors want to thank F. Gelis, R. Venugopalan, T. Lappi, Y. Kovchegov, K. Tuchin and J.-S. Gagnon for interesting and stimulating discussions. 

\end{acknowledgments}


\appendix

\section{Cross section in $pp$ collisions}
\label{app:pp_k_perp}

The calculation of the cross section for $\eta'$ production in proton-proton collisions at RHIC can be performed in the $k_{\perp}$-factorization formalism. This formalism can be used in the semihard regime where the factorization scale satisfies the inequality $\Lambda_{\rm{QCD}}^{2} \ll Q^{2} \ll \sqrt{s}$. For $\eta'$, this inequality is only marginally satisfied at small transverse momentum because its mass is relatively low. It is recovered at larger transverse momentum, around $|k_{\perp}| \sim 2 - 3 \; \mbox{GeV}$. The starting point is the usual formula of the inclusive cross section in $k_{\perp}$-factorization:
\begin{eqnarray}
(2\pi)^{3}2E_{k}\frac{d\sigma^{pp \rightarrow \eta' X}}{d^{3}k} &=& 16 \pi^{2} \int_{0}^{1}\frac{dx_{1}}{x_{1}}\frac{dx_{2}}{x_{2}}\int \frac{d^{2}q_{\perp} d^{2}p_{\perp}}{(2\pi)^{4}} \phi_{1}(x_{1},p_{\perp}^{2},Q^2) \phi_{2}(x_{2},q_{\perp}^{2},Q^2) \nonumber \\
&& \times (2\pi)^{3}2E_{k}\frac{d\sigma^{g^{*}g^{*} \rightarrow \eta' X}}{d^{3}k}
\end{eqnarray}
where
\begin{eqnarray}
(2\pi)^{3}2E_{k}\frac{d\sigma^{g^{*}g^{*} \rightarrow H}}{d^{3}k} = \frac{1}{2\hat{s}} |\mathcal{M}^{g^{*}g^{*} \rightarrow \eta'}|^{2} (2\pi)^{4} \delta^{4}(p+q-k)
\end{eqnarray}
is the high-energy limit of the cross section for off-shell gluons $g^{*}$ to on-shell $\eta'$ mesons, $\hat{s} = x_{1}x_{2}s$ is the $k_{\perp}$-factorization flux factor \cite{Collins:1991ty,Catani:1990eg}, $x_{1,2}$ are momentum fractions of gluons and $\phi_{1,2}(x_{1,2},k_{\perp},Q^2)$ are unintegrated gluon distribution functions of proton 1 and 2. The unintegrated distribution functions are related to usual parton distribution functions of gluons (appearing in collinear factorization ) by
\begin{eqnarray}
\label{eq:k_vs_coll}
\int_{0}^{Q^2} dk_{\perp}^{2} \phi(x,k_{\perp}^{2},Q^2) \approx xG(x,Q^2)
\end{eqnarray}
where $G(x,Q^2)$ is the usual gluon distribution function in collinear factorization.

To compute the production cross section of $\eta'$ mesons, the high energy limit of the lowest order matrix element between two off-shell gluons and one on-shell $\eta'$ has to be calculated. This can be evaluated by using Feynman rules where the vertex, evaluated from the interaction Lagrangian, is given by Eq. (\ref{eq:vertex}). Note also that in $k_{\perp}$-factorization, the sum on polarization tensors is given by \cite{Collins:1991ty,Catani:1990eg}  
\begin{eqnarray}
\sum_{\lambda} \epsilon^{*\mu}_{\lambda}(p) \epsilon^{\nu}_{\lambda}(p) = \frac{p^{\mu}_{\perp}p^{\nu}_{\perp}}{p_{\perp}^{2}}
\end{eqnarray}
where $p_{\perp}^{\mu} \equiv (0,p_{\perp},0)$. The sum on polarizations differs from the usual result because we are considering off-shell gluons with a virtuality given by $p^{2}=-p_{\perp}^{2}$. The exact form is due to the coupling of gluons to partons through eikonal vertices as well as gauge invariance and Ward identities \cite{Catani:1990eg}.

In the center of mass frame, the 4-momenta of partons inside the proton moving in the $\pm z$ direction in Minkowski coordinates can be written,as:
\begin{eqnarray}
P &=& \left(\frac{\sqrt{s}}{2},0,0, \frac{\sqrt{s}}{2}\right) \; ; \;
Q = \left(\frac{\sqrt{s}}{2},0,0, -\frac{\sqrt{s}}{2}\right)
\end{eqnarray}
Then, the momenta of gluons in the large energy limit ($|p_{\perp}|,|q_{\perp}| \ll \sqrt{s}$) are simply 
\begin{eqnarray}
\label{eq:momenta}
p &=& \left(\frac{x_{1}\sqrt{s}}{2},p_{\perp}, \frac{x_{1}\sqrt{s}}{2}\right) \; ; \;
q = \left(\frac{x_{2}\sqrt{s}}{2},q_{\perp}, -\frac{x_{2}\sqrt{s}}{2}\right)
\end{eqnarray}
Using this kinematics, it is possible to compute the high-energy limit of the matrix element $\mathcal{M}$. This can then be inserted in the expression of the cross section which is finally given by (see \cite{Lipatov:2005at,Luszczak:2005xs,FillionGourdeau:2007ee} for more details and similar calculations)
\begin{eqnarray}
\label{eq:k_fact_cross}
(2\pi)^{3}2E_{k}\frac{d\sigma^{pp \rightarrow \eta' X}}{d^{3}k} &=& \frac{4\pi^{2}}{(N_{c}^{2} -1)} \int d^{2}p_{\perp}d^{2}q_{\perp} \phi_{1}(x_{+},p_{\perp}^{2},Q^2) \phi_{2}(x_{-},q_{\perp}^{2},Q^2) \nonumber \\
&& \times \delta^{2}(k_{\perp} - p_{\perp} - q_{\perp}) |F(-p_{\perp}^{2},-q_{\perp}^{2})|^{2} \frac{\left[\epsilon_{ij} p^{i} q^{j} \right]^{2}}{p_{\perp}^{2} q_{\perp}^{2}}
\end{eqnarray}
where $x_{\pm} = \frac{M_{\perp}}{\sqrt{s}} e^{\pm y}$. This expression can be used to study the phenomenology of $\eta'$ production in $pp$ collisions.

\subsection{Limit of Collinear Factorization}

The procedure to recover collinear factorization cross sections from $k_{\perp}$-factorization is well-known \cite{Collins:1991ty,Lipatov:2001ny,Zotov:2003cb,Gelis:2003vh} and will serve as a consistency check for Eq. (\ref{eq:k_fact_cross}). The limit $|p_{\perp}|,|q_{\perp}| \rightarrow 0$ has to be taken in the matrix elements and the integration on the azimuthal angle has to be performed. The last step is to use the relation Eq. (\ref{eq:k_vs_coll}) to make the last integral and relate the unintegrated distributions to the collinear distributions.  We obtain
\begin{eqnarray}
\label{eq:k_coll_cross}
(2\pi)^{3}2E_{k}\frac{d\sigma_{\rm{coll.}}^{pp \rightarrow \eta' X}}{d^{3}k} &=& \frac{\pi^{2} M^{2} |F(0,0)|^{2}}{2s(N_{c}^{2} -1)} G_{1}(x'_{+},Q^2) G_{2}(x'_{-},Q^2) (2\pi)^{2} \delta^{2}(k_{\perp}) 
\end{eqnarray}
where $x'_{\pm} = \frac{M}{\sqrt{s}} e^{\pm y}$. This expression corresponds exactly to the well-known result for leading-order $\eta'$ production in pQCD collinear formalism \cite{JalilianMarian:2001bu}.


\section{Simplification of Eq. (\ref{eq:simp_cross})}
\label{app:simp}

In this Appendix, we simplify the second term of the cross section and put it in a form that can be evaluated numerically. First, using the definition of $\bar{V}$ given by Eq. (\ref{eq:two_av}), we can perform the integrals in the exponent and we get
\begin{eqnarray}
\bar{V}(z_{\perp},0) &=&  \exp \biggl\{  -Q_{s}^{2} \biggl[ \frac{z_{\perp}^{2}}{4} \left( 1-\gamma  + \ln \left[ \frac{2}{\Lambda_{\rm{QCD}} |z_{\perp}|}  \right] \right) \nonumber \\
&& + \frac{z_{\perp}^{4} \Lambda_{\rm{QCD}}^{2}}{128} \;_{2}F_{3}\left(1,1;2,3,3 \left| -\frac{z_{\perp}^{2} \Lambda^{2}_{\rm{QCD}}}{4} \right.\right) \biggr] \biggr\}
\end{eqnarray}
where $\gamma$ is Euler constant and where $\;_{2}F_{3}\left(1,1;2,3,3 \left| z \right.\right)$ is the generalized hypergeometric series. The constant $\Lambda_{\rm{QCD}}$ appears in this expression as an infrared regulator.  

Then, we consider the second term of Eq. (\ref{eq:simp_cross}) given by
\begin{eqnarray}
(2\pi)^3 2E_k \frac{d\sigma_{2}}{ d^3 k}
&=&
\pi^{2}g^{2} \mu^{2}_{A} S_{\perp}  \int d^{2}z_{\perp} \frac{d^{2}p_{\perp} d^{2}r_{\perp} }{(2\pi)^{4}} |F(-p_{\perp}^{2} , -p_{2,\perp}^{2})|^{2} \epsilon_{ij}\epsilon_{kl} \frac{p^{i} p^{k} r^{j} r^{l}}{p^{2}_{\perp} (p^{2}_{\perp}+ \Lambda^{2}) r_{\perp}^{2} }     \nonumber \\
&& \times  \phi_{p}(r_{\perp}^{2},x)         e^{i( p_{2,\perp} - r_{\perp}) \cdot z_{\perp}} \bar{V}(z_{\perp},0) .
\end{eqnarray}  
By letting $v_{\perp} = k_{\perp} - p_{\perp} - r_{\perp}$ and by using the identity
\begin{eqnarray}
\int_{0}^{2\pi} d\theta e^{ix\cos(\theta)} &=& 2\pi J_{0}(x)
\end{eqnarray}
(where $J_{0}(x)$ is a Bessel function of the first kind), we find that
\begin{eqnarray}
(2\pi)^3 2E_k \frac{d\sigma_{2}}{d^3 k}
&=&
2\pi^{3}g^{2} \mu^{2}_{A} S_{\perp} \int \frac{d^{2}p_{\perp} d^{2}v_{\perp} }{(2\pi)^{4}} |F(-p_{\perp}^{2} , -p_{2,\perp}^{2})|^{2} \epsilon_{ij}\epsilon_{kl} \frac{p^{i} p^{k} (k-p-v)^{j} (k-p-v)^{l}}{p^{2}_{\perp} (p^{2}_{\perp}+ \Lambda^{2}) (k-p-v)_{\perp}^{2} }     \nonumber \\
&& \times  \phi_{p}[(k-p-v)_{\perp}^{2},x]  \Gamma(|v_{\perp}|) .
\end{eqnarray}  
where
\begin{eqnarray}
\Gamma(|v_{\perp}|) \equiv \int_{0}^{\infty} dz zJ_{0}(|v_{\perp}|z)\bar{V}(z).
\end{eqnarray}
We let $q_{\perp} = p_{\perp} + v_{\perp}$ and we use polar coordinates to get
\begin{eqnarray}
\label{eq:final_simp}
(2\pi)^3 2E_k \frac{d\sigma_{2}}{ d^3 k}
&=&
\frac{\alpha_{s}(M_{\perp}^{2}) \mu^{2}_{A} S_{\perp}}{2} \int_{0}^{\infty}dpdq \int_{0}^{2\pi} d\theta d\Theta  |F(-p^{2} , -p^{2}-k^{2}+2pk\cos(\theta))|^{2} \nonumber \\
&& \times  pq\frac{[q\sin(\theta - \Theta) - k\sin(\theta)]^{2}}{(p^{2}+\Lambda^{2})[q^{2}+k^{2}-2qk\cos(\Theta)]}   \nonumber \\
&& \times  \phi_{p}[q^{2}+k^{2}-2qk\cos(\Theta),x]  \Gamma(|q^{2}+p^{2}-2pq\cos(\theta - \Theta)|) .
\end{eqnarray}  
where $\alpha_{s}$ is the strong coupling constant. This expression is integrated numerically and the result is shown in section \ref{sec:num_res}.

\section{Computation of Eq. (\ref{eq:cross01}) in covariant gauge}
\label{app:cov}

In this Appendix, we compute the cross section by using the solution of the Yang-Mills equation in covariant gauge. This serves as a check for Eq. (\ref{eq:cross01}) and it shows that our result is gauge independent. The solution of the gauge field is given by $A^{\mu}_{a}(k) = A^{\mu}_{p , a}(k) + A^{\mu}_{A,a}(k) + A^{\mu}_{pA,a}$ where $A^{\mu}_{p,a}(k)$ is the field associated with the proton (of $O(\rho_{p})$), $A^{\mu}_{A, a}(k)$ is the field associated with the nuclei (of $O(\rho_{A}) \sim O(1)$) and $A^{\mu}_{pA,a}(k)$ is the field produced by the collision (of $O(\rho_{p}\rho_{A}^{\infty}) \sim O(\rho_{p})$) \cite{Blaizot:2004wu,Kovchegov:1998bi}. The fields $A^{\mu}_{p}$ and $A^{\mu}_{pA}$ are weak and are used as small parameters to solve the Yang-Mills equation perturbatively. The field $A^{\mu}_{A}$ is strong and satisfies $A^{\mu}_{p},A^{\mu}_{pA} \ll A^{\mu}_{A}$. The explicit solution is given by \cite{Blaizot:2004wu,Kovchegov:1998bi}
\begin{eqnarray}
\label{eq:gauge_p_cov}
A^{+}_{p ,a}(k) &=&  2\pi g \delta(k^{-}) \frac{\rho_{p,a}(k_{\perp})}{k_{\perp}^{2}}  \\
\label{eq:gauge_A_cov}
A^{-}_{A,a}(k) &=& 2\pi g \delta(k^{+}) \frac{\rho_{A,a}(k_{\perp})}{k_{\perp}^{2}} \\
\label{eq:gauge_prod_cov}
A^{i}_{pA,a}(k) &=& -\frac{ig}{k^{2} + ik^{+} \epsilon} \int \frac{d^{2}p_{\perp}}{(2\pi)^{2}}\frac{\rho_{p,b}(p_{\perp})}{p_{\perp}^{2}}  \nonumber \\
&& \times \biggl\{  C_{U}^{\mu}(k,p_{\perp}) \left[ U_{ab}(k_{\perp} - p_{\perp}) - (2\pi)^{2} \delta^{2}(k_{\perp} - p_{\perp}) \delta_{ab}   \right] \nonumber \\
&& +  C_{V}^{\mu}(k,p_{\perp})\left[ V_{ab}(k_{\perp} - p_{\perp}) - (2\pi)^{2} \delta^{2}(k_{\perp} - p_{\perp}) \delta_{ab}   \right] \biggr\}
\end{eqnarray} 
where $V_{ab}(k_{\perp} - p_{\perp})$ is a Wilson line with a $\frac{1}{2}$ coefficient in the exponential (see \cite{Blaizot:2004wu}) and 
\begin{eqnarray}
&&C_{U}^{+}(k,p_{\perp}) \equiv - \frac{p_{\perp}^{2}}{k^{-} + i\epsilon} , C_{U}^{-}(k,p_{\perp}) \equiv  \frac{p_{\perp}^{2}-2p_{\perp} \cdot k_{\perp}}{k^{+} } , C_{U}^{i}(k,p_{\perp}) \equiv -2p^{i}, \nonumber \\
&& C_{V}^{+}(k,p_{\perp}) \equiv 2k^{+} , C_{V}^{-}(k,p_{\perp}) \equiv 2\frac{k_{\perp}^{2}}{k^{+}} ,  C_{V}^{i}(k,p_{\perp}) \equiv 2k^{i}.
\end{eqnarray}
The power counting is very similar to the light-cone gauge. We get that the leading order contribution to the field-strength correlator is given by
\begin{eqnarray}
B_{z,z'}^{\rm{cov}}(k)&=&64 \int \frac{d^{4}p d^{4}q}{(2\pi)^{8}}  F^{*}(-p_{\perp}^{2} , -p_{2,\perp}^{2})  F(-q_{\perp}^{2},-q_{2,\perp}^{2})  \epsilon_{ij}\epsilon_{kl} \nonumber \\
&& \times \langle \left[ q^{i}(k-q)^{j}A_{A,b}^{-} (q) A_{z,b}^{+ }(k-q) - q^{i}(k-q)^{+} A_{A,b}^{-} (q) A_{z,b}^{j}(k-q) \right] \nonumber \\
&& \times  \bigl[ p^{k}(k-p)^{l}A_{A,a}^{- *} (p) A_{z',a}^{+ *}(k-p) - p^{k}(k-p)^{+} A_{A,a}^{- *} (p) A_{z',a}^{l *}(k-p) \bigr]    \rangle .
\end{eqnarray}
Using the expression of the gauge field, it is a straightforward calculation to show that $B_{p,p}^{LC}(k)=B_{p,p}(k)$. Just like in light-cone gauge, the other terms require more work. We will first look at
\begin{eqnarray}
B_{pA,p}^{\rm{cov}}(k)&=&64g^{3} \int \frac{d^{2}p_{\perp} d^{2}q_{\perp}dq^{-}}{(2\pi)^{5}}  F^{*}(-p_{\perp}^{2} , -p_{2,\perp}^{2})  F(-q_{\perp}^{2},-q_{2,\perp}^{2})  \epsilon_{ij}\epsilon_{kl} \frac{p^{k}(k-p)^{l}}{q_{\perp}^{2} p_{\perp}^{2}(k-p)_{\perp}^{2}} \nonumber \\
&& \times \langle \left[ q^{i}(k-q)^{j} A_{pA,b}^{+ }(k-q) - q^{i}k^{+}  A_{pA,b}^{j}(k-q) \right] \nonumber \\
&& \times  \rho_{A,b}(q_{\perp}) \rho_{A,a}^{*}(p_{\perp}) \rho_{p,a}^{*}(k_{\perp} - p_{\perp})    \rangle 
\end{eqnarray}
which is obtained by replacing the gauge fields with their explicit expression. Now, we have that
\begin{eqnarray}
&&q^{i}(k-q)^{j} A_{pA,b}^{+ }(k-q) - q^{i}k^{+}  A_{pA,b}^{j}(k-q) = \nonumber \\
&& \frac{ig}{(k-q)^{2} + ik^{+} \epsilon} \int \frac{d^{2}r_{\perp}}{(2\pi)^{2}} \frac{\rho_{p,e}(r_{\perp})}{r_{\perp}^{2}} \biggl[ \frac{q^{i} (k-q)^{j} r_{\perp}^{2}}{k^{-} - q^{-} + i\epsilon} + 2q^{i}k^{+}r^{j}  \biggr] \nonumber \\
&& \times \left[ U_{be}(k_{\perp} - q_{\perp} - r_{\perp}) - (2\pi)^{2} \delta^{2}(k_{\perp} - q_{\perp} - r_{\perp}) \delta_{be}   \right] 
\end{eqnarray}
which does not depend on $V_{ab}(k_{\perp})$. This is required because this Wilson line does not appear in the light-cone gauge calculation. Furthermore, the first term is zero when it is integrated on $q^{-}$ because it has two poles on the side of the real axis (similar to the LC calculation in section \ref{sec:calc_corr}). The integration on $q^{-}$ of the second term can be done with the principal part identity (see again section \ref{sec:calc_corr}). The result is that $B_{pA,p}^{\rm{cov}}(k) =B_{pA,p}(k) $. A similar calculation can be performed with the two other terms and we find that $B^{\rm{cov}}(k)=B(k)$.

\section{Diagrammatic rules and $n$ Wilson lines - $m$ color charges densities correlators}
\label{app:dia_rules}

In this appendix, we discuss the diagrammatic rules used in Sec. \ref{sec:averages} to compute the Wilson lines - color charge densities correlators. These techniques can be used in any representation of the $SU(N)$ generators, as long as the Wilson line correlators such as
\begin{eqnarray}
\label{eq:wilson_line_corr}
F^{j}(b^{+},a^{+}|\{a\},\{b\}) \equiv \langle U_{a_{1}b_{1}}(b^{+},a^{+}|x_{1\perp})U_{a_{2}b_{2}}(b^{+},a^{+}|x_{2\perp})...U_{a_{j}b_{j}}(b^{+},a^{+}|x_{j\perp})\rangle
\end{eqnarray}
are known, which we assume throughout the following discussion. Here, $\{a\},\{b\}$ are the sets of color indices defined as $\{a_{1},a_{2},...,a_{j}\}$ and $\{b_{1},b_{2},...,b_{j}\}$ respectively. The most general correlator we study here is
\begin{eqnarray}
F^{m,n} (b^{+},a^{+})&\equiv& \langle  \rho_{c_{1}}(y^{+}_{1},y_{1\perp}) \rho_{c_{2}}(y^{+}_{2},y_{2\perp})...\rho_{c_{m}}(y^{+}_{m},y_{m\perp}) \nonumber \\
&&\; \; \; \; \; \; \; \; \;\times U_{a_{1}b_{1}}(b^{+},a^{+}|x_{1\perp})U_{a_{2}b_{2}}(b^{+},a^{+}|x_{2\perp})...U_{a_{n}b_{n}}(b^{+},a^{+}|x_{n\perp})\rangle .
\end{eqnarray}
We want to express this correlator in terms of Wilson line correlators shown in Eq. (\ref{eq:wilson_line_corr}). This can be done in a general way. First, we start by defining a number of new quantities necessary for the computation. We define a quantity that represents a correlator with a number $j$ of color charge densities such as
\begin{eqnarray}
G^{j}_{(1,2,...,\{k,k+1\},...,j+1,j+2)} &\equiv & \langle  \rho_{c_{1}}(y_{1}^{+},y_{1\perp}) \rho_{c_{2}}(y_{2}^{+},y_{2\perp})... \rho_{c_{k-1}}(y^{+}_{k-1},y_{k-1\perp}) \rho_{c_{k+2}}(y^{+}_{k+2},y_{k+2\perp}) ... \nonumber \\
&& \times \rho_{c_{j+2}}(y^{+}_{j+2},y_{j+2\perp}) \rangle .
\end{eqnarray}
In this definition of $G^{j}_{(1,2,...,\{k,k+1\},...,j+1,j+2)}$, the upper index counts the number of sources while the lower index indicates in bracket which sources are missing. We also define $H^{j,m}$ in a similar way by
\begin{eqnarray}
\label{eq:H_def}
H^{j,n}_{(1,2,...,\{k,k+1\},...,j+1,j+2)} &=& \langle  \rho_{c_{1}}(y_{1}^{+},y_{1\perp}) \rho_{c_{2}}(y_{2}^{+},y_{2\perp})... \rho_{c_{k-1}}(y^{+}_{k-1},y_{k-1\perp}) \rho_{c_{k+2}}(y^{+}_{k+2},y_{k+2\perp}) ... \nonumber \\
&& \times \rho_{c_{j+2}}(y^{+}_{j+2},y_{j+2\perp}) \nonumber \\
&&\times U_{a_{1}b_{1}}(b^{+},a^{+}|x_{1\perp})U_{a_{2}b_{2}}(b^{+},a^{+}|x_{2\perp})...U_{a_{n}b_{n}}(b^{+},a^{+}|x_{n\perp})\rangle_{\mathrm{conn.}} 
\end{eqnarray}
where $j \leq m$ and where the subscript $\mathrm{conn.}$ indicates that only the connected part of the correlator is considered, which means that all \textit{external} sources are contracted with an \textit{internal} source (note that \textit{external} and \textit{internal} are defined in Sec. \ref{subsec:1W1cav}). Now, there are two possible cases: $m$ can be odd or even.

\begin{enumerate}
\item Even $m$

For the case where $m$ is even, we can use Wick theorem to write
\begin{eqnarray}
F^{m,n}(b^{+},a^{+}) &\equiv&  G^{m} H^{0,n} +  \sum_{i,j,i<j} G^{m-2}_{(1,...,i-1,\{i\},i+1,...,j-1,\{j\},j+1,...,m)} H^{2,n}_{(\{1,...,i-1\},i,\{i+1,...,j-1\},j,\{j+1,...,m\})} \nonumber \\
&& + \sum_{i,j,k,l,i<j<k < l} G^{m-4}_{(1,...,i-1,\{i\},i+1,...,j-1,\{j\},j+1,...,k-1,\{k\},k+1,...,l-1,\{l\},l+1,...,m)} \nonumber \\
&& \;\;\;\;\;\;\;\;\;\;\;\;\;\;\;\;\;\;\;\; \times  H^{4,n}_{(\{1,...,i-1\},i,\{i+1,...,j-1\},j,\{j-1,...,k-1\},k,\{k-1,...,l-1\},l,\{l+1,...,m\})}\nonumber \\
&&+ ... +  \sum_{i,j,i<j} G^{2}_{(\{1,...,i-1\},i,\{i+1,...,j-1\},j,\{j+1,...,m\})} H^{2,n}_{(1,...,i-1,\{i\},i+1,...,j-1,\{j\},j+1,...,m)} \nonumber \\
&& + H^{m,n} .
\end{eqnarray}

\item Odd $m$

For the case where $m$ is odd, we can use Wick theorem to write
\begin{eqnarray}
F^{m,n}(b^{+},a^{+}) &\equiv&    \sum_{i} G^{m-1}_{(1,...,i-1,\{i\},i+1,...,m)} H^{2,n}_{(\{1,...,i-1\},i,\{i+1,...,m\})} \nonumber \\
&& + \sum_{i,j,k,i<j<k } G^{m-3}_{(1,...,i-1,\{i\},i+1,...,j-1,\{j\},j+1,...,k-1,\{k\},k+1,...,m)} \nonumber \\
&& \;\;\;\;\;\;\;\;\;\;\;\;\;\;\;\;\;\;\;\; \times  H^{3,n}_{(\{1,...,i-1\},i,\{i+1,...,j-1\},j,\{j-1,...,k-1\},k,\{k-1,...,m\})}\nonumber \\
&&+ ... + \sum_{i,j,i<j} G^{2}_{(\{1,...,i-1\},i,\{i+1,...,j-1\},j,\{j+1,...,m\})} H^{2,n}_{(1,...,i-1,\{i\},i+1,...,j-1,\{j\},j+1,...,m)} \nonumber \\
&& + H^{m,n}.
\end{eqnarray}

\end{enumerate}

\begin{figure}
\centering
		\includegraphics[width=0.7\textwidth]{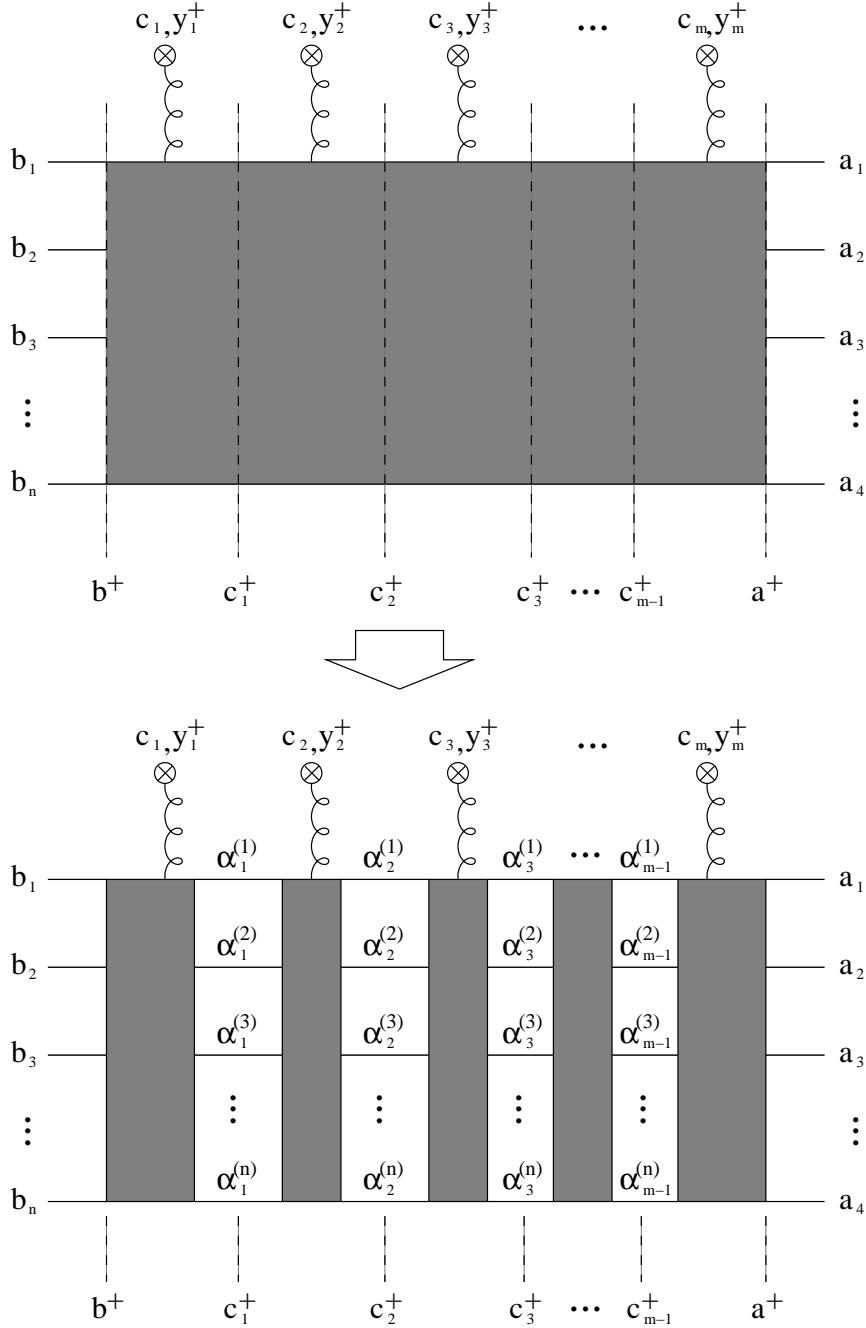}
	\caption{Splitting the connected correlator $H^{m,n} (b^{+},a^{+}|\{a\},\{b\})$, shown in the first diagram, into different slices according to Eq. (\ref{eq:Hmn}). Here, $a_{j},b_{j},c_{j},\alpha_{j}^{i}$ are color indices and $a^{+},b^{+},c_{j}^{+},y_{j}^{+}$ are variables in the $+$ coordinate. }
	\label{fig:gen_corr}
\end{figure}

\begin{figure}
\centering
		\includegraphics[width=0.9\textwidth]{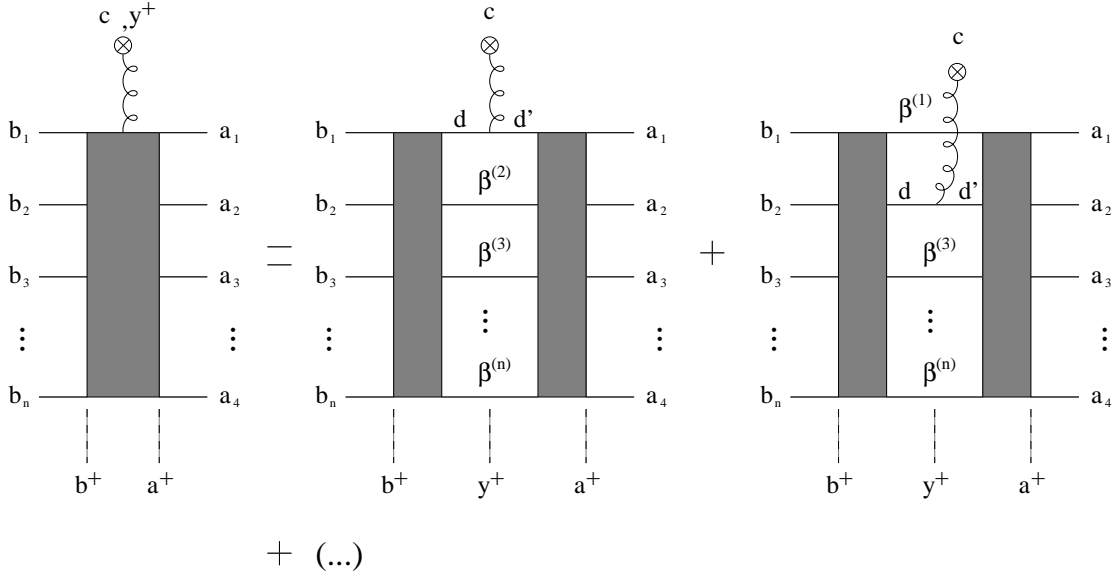}
	\caption{Inserting a source to compute $H^{1,n} (b^{+},a^{+}|\{a\},\{b\})$, which is shown on the left diagrams. The (...) contains all the other ways of inserting the source on the other branches. $\beta^{i}$ are color indices.}
	\label{fig:one_source_corr}
\end{figure}
Then, we have to evaluate explicitly the connected part of the correlator $H^{j,n}$. For the following, we use the definition 
\begin{eqnarray}
\label{eq:Hmn_def}
H^{m,n} (b^{+},a^{+}|\{a\},\{b\})&\equiv& \langle  \rho_{c_{1}}(y^{+}_{1},y_{1\perp}) \rho_{c_{2}}(y^{+}_{2},y_{2\perp})...\rho_{c_{m}}(y^{+}_{m},y_{m\perp}) \nonumber \\
&&\times U_{a_{1}b_{1}}(b^{+},a^{+}|x_{1\perp})U_{a_{2}b_{2}}(b^{+},a^{+}|x_{2\perp})...U_{a_{n}b_{n}}(b^{+},a^{+}|x_{n\perp})\rangle_{\mathrm{conn.}} 
\end{eqnarray}
where we assume for simplicity that $b^{+} > y_{1}^{+} > y_{2}^{+}> ... > y_{m-1}^{+} > y_{m}^{+} > a^{+}$. To get the general result where all the $y^{+}_{j}$ can have any value between $b^{+}$ and $a^{+}$, one has to sum over all orderings. All other cases where a source is missing such as in the definition Eq. (\ref{eq:H_def}) can be treated in a similar way. It is now convenient to define a new set of variables $c_{j}^{+}$ which have the following property $y_{j}^{+}>c_{j}^{+} >  y_{j+1}^{+}$ . Then, we use these new variables to split the $+$ coordinate in slices, which allow us to write the Wilson lines (using their properties) as 
\begin{eqnarray}
\label{eq:U_split}
U_{a_{j}b_{j}}(b^{+},a^{+}|x_{j\perp}) = U_{a_{j}\alpha_{1}^{(j)}}(b^{+},c_{1}^{+}|x_{j\perp}) \left[ \prod_{p=1}^{m-2} U_{\alpha_{p}^{(j)}\alpha^{(j)}_{p+1}}(c_{p}^{+},c_{p+1}^{+}|x_{j\perp}) \right] U_{\alpha^{(j)}_{m+1}b_{j}}(c^{+}_{m-1},a^{+}|x_{j\perp}).
\end{eqnarray}
Color charge densities in different slices cannot be contracted using Wick theorem because such contractions do not have support. Using this fact and Eq. (\ref{eq:U_split}), we get that
\begin{eqnarray}
\label{eq:Hmn}
H^{m,n} (b^{+},a^{+}|\{a\},\{b\})&\equiv& H^{1,n}(b^{+},c_{1}^{+}|\{a\},\{\alpha_{1}\}) \left[\prod_{p=1}^{m-2} H^{1,n}(c_{p}^{+},c_{p+1}^{+}|\{\alpha_{p}\},\{\alpha_{p+1}\})  \right] \nonumber \\
&& \times H^{1,n}(c_{m-1}^{+},a^{+}|\{\alpha_{m-1}\},\{a\}).
\end{eqnarray}
Eqs. (\ref{eq:Hmn_def}),(\ref{eq:Hmn}) and the process of slicing the $+$ coordinate are shown diagrammatically in Fig. \ref{fig:gen_corr}.

The only unknown quantity left is $ H^{1,n}(b^{+},a^{+}|\{a\},\{b\})$, to which we now turn. Remember that in this quantity, there is only one \textit{external} source and this source, using Wick theorem, can be contracted with an \textit{internal} source in any of the $n$ Wilson line included in $ H^{1,n}(b^{+},a^{+}|\{a\},\{b\})$. In Sec. \ref{subsec:1W1cav}, we show that when a source is contracted with a Wilson line, it breaks the Wilson line in two, change the color structure and multiply the overall expression by $\mu^{2}(y^{+})G_{0}(x_{\perp})$. For the general case, it is shown in Fig. (\ref{fig:one_source_corr}). Using this and the definition of $H$, we get 
\begin{eqnarray}
H^{1,n}(b^{+},c_{1}^{+}|\{a\},\{b\}) &=& \sum_{j=1}^{n} \mu^{2}(y^{+}) G_{0}(y_{\perp} - x_{j \perp}) f_{cdd'} \nonumber \\
&& \times \left. F^{n}(b^{+},y^{+}|\{a\},\{ \beta\})\right|_{\beta_{j} = d} \left.F^{n}(y^{+},a^{+}|\{ \beta \},\{ b \})\right|_{\beta_{j} = d'}.
\end{eqnarray}
which completes our calculation since the $F^{n}$'s are assumed to be known from the beginning. 

This general result can be summarized into diagrammatic rules to compute any $F^{m,n}$. 
\begin{itemize}
\item Determine a $+$-ordering for the \textit{external} sources.
\item Draw all possible ways the \textit{external} sources can be contracted with the Wilson lines and with each other by respecting the $+$-ordering.
\item A blob with $n$ external line is given by $F^{j}(b^{+},a^{+}|\{a\},\{b\})$.
\item A source insertion at $y^{+}$ with a color index $c$ on line $j$ multiplies the expression by $\mu^{2}(y^{+})G_{0}(y_{\perp} - x_{j\perp})f_{cdd'}$ and changes the indices on the left blobs to $\beta^{(j)}=d$ and on the right blob to $\beta^{(j)}=d'$.
\item \textit{External} sources that are contracted together gives $\mu^{2}(y_{1}^{+}) \delta_{c_{1}c_{2}} \delta(y_{1}^{+} - y_{2}^{+})\delta^{2}(y_{1 \perp} - y_{2 \perp})$.
\item Sum on all $+$-orderings.
\end{itemize}
Examples of these rules are shown in Sec. \ref{sec:averages}.

\bibliographystyle{apsrev}
\bibliography{bibliography}

\begin{thebibliography}{68}
\expandafter\ifx\csname natexlab\endcsname\relax\def\natexlab#1{#1}\fi
\expandafter\ifx\csname bibnamefont\endcsname\relax
  \def\bibnamefont#1{#1}\fi
\expandafter\ifx\csname bibfnamefont\endcsname\relax
  \def\bibfnamefont#1{#1}\fi
\expandafter\ifx\csname citenamefont\endcsname\relax
  \def\citenamefont#1{#1}\fi
\expandafter\ifx\csname url\endcsname\relax
  \def\url#1{\texttt{#1}}\fi
\expandafter\ifx\csname urlprefix\endcsname\relax\def\urlprefix{URL }\fi
\providecommand{\bibinfo}[2]{#2}
\providecommand{\eprint}[2][]{\url{#2}}

\bibitem[{\citenamefont{Collins and Ellis}(1991)}]{Collins:1991ty}
\bibinfo{author}{\bibfnamefont{J.~C.} \bibnamefont{Collins}} \bibnamefont{and}
  \bibinfo{author}{\bibfnamefont{R.~K.} \bibnamefont{Ellis}},
  \bibinfo{journal}{Nucl. Phys.} \textbf{\bibinfo{volume}{B360}},
  \bibinfo{pages}{3} (\bibinfo{year}{1991}).

\bibitem[{\citenamefont{Catani et~al.}(1991)\citenamefont{Catani, Ciafaloni,
  and Hautmann}}]{Catani:1990eg}
\bibinfo{author}{\bibfnamefont{S.}~\bibnamefont{Catani}},
  \bibinfo{author}{\bibfnamefont{M.}~\bibnamefont{Ciafaloni}},
  \bibnamefont{and} \bibinfo{author}{\bibfnamefont{F.}~\bibnamefont{Hautmann}},
  \bibinfo{journal}{Nucl. Phys.} \textbf{\bibinfo{volume}{B366}},
  \bibinfo{pages}{135} (\bibinfo{year}{1991}).

\bibitem[{\citenamefont{Gribov et~al.}(1983)\citenamefont{Gribov, Levin, and
  Ryskin}}]{Gribov:1984tu}
\bibinfo{author}{\bibfnamefont{L.~V.} \bibnamefont{Gribov}},
  \bibinfo{author}{\bibfnamefont{E.~M.} \bibnamefont{Levin}}, \bibnamefont{and}
  \bibinfo{author}{\bibfnamefont{M.~G.} \bibnamefont{Ryskin}},
  \bibinfo{journal}{Phys. Rept.} \textbf{\bibinfo{volume}{100}},
  \bibinfo{pages}{1} (\bibinfo{year}{1983}).

\bibitem[{\citenamefont{Kuraev et~al.}(1977)\citenamefont{Kuraev, Lipatov, and
  Fadin}}]{Kuraev:1977fs}
\bibinfo{author}{\bibfnamefont{E.~A.} \bibnamefont{Kuraev}},
  \bibinfo{author}{\bibfnamefont{L.~N.} \bibnamefont{Lipatov}},
  \bibnamefont{and} \bibinfo{author}{\bibfnamefont{V.~S.} \bibnamefont{Fadin}},
  \bibinfo{journal}{Sov. Phys. JETP} \textbf{\bibinfo{volume}{45}},
  \bibinfo{pages}{199} (\bibinfo{year}{1977}).

\bibitem[{\citenamefont{Luszczak and
  Szczurek}(2006{\natexlab{a}})}]{Luszczak:2005cq}
\bibinfo{author}{\bibfnamefont{M.}~\bibnamefont{Luszczak}} \bibnamefont{and}
  \bibinfo{author}{\bibfnamefont{A.}~\bibnamefont{Szczurek}},
  \bibinfo{journal}{Phys. Rev.} \textbf{\bibinfo{volume}{D73}},
  \bibinfo{pages}{054028} (\bibinfo{year}{2006}{\natexlab{a}}),
  \eprint{hep-ph/0512120}.

\bibitem[{\citenamefont{Lipatov et~al.}(2001)\citenamefont{Lipatov, Saleev, and
  Zotov}}]{Lipatov:2001ny}
\bibinfo{author}{\bibfnamefont{A.~V.} \bibnamefont{Lipatov}},
  \bibinfo{author}{\bibfnamefont{V.~A.} \bibnamefont{Saleev}},
  \bibnamefont{and} \bibinfo{author}{\bibfnamefont{N.~P.} \bibnamefont{Zotov}}
  (\bibinfo{year}{2001}), \eprint{hep-ph/0112114}.

\bibitem[{\citenamefont{Zotov et~al.}(2003)\citenamefont{Zotov, Lipatov, and
  Saleev}}]{Zotov:2003cb}
\bibinfo{author}{\bibfnamefont{N.~P.} \bibnamefont{Zotov}},
  \bibinfo{author}{\bibfnamefont{A.~V.} \bibnamefont{Lipatov}},
  \bibnamefont{and} \bibinfo{author}{\bibfnamefont{V.~A.}
  \bibnamefont{Saleev}}, \bibinfo{journal}{Phys. Atom. Nucl.}
  \textbf{\bibinfo{volume}{66}}, \bibinfo{pages}{755} (\bibinfo{year}{2003}).

\bibitem[{\citenamefont{Jung}(2002)}]{Jung:2001rp}
\bibinfo{author}{\bibfnamefont{H.}~\bibnamefont{Jung}}, \bibinfo{journal}{Phys.
  Rev.} \textbf{\bibinfo{volume}{D65}}, \bibinfo{pages}{034015}
  (\bibinfo{year}{2002}), \eprint{hep-ph/0110034}.

\bibitem[{\citenamefont{Andersson et~al.}(2002)}]{Andersson:2002cf}
\bibinfo{author}{\bibfnamefont{B.}~\bibnamefont{Andersson}}
  \bibnamefont{et~al.} (\bibinfo{collaboration}{Small x}),
  \bibinfo{journal}{Eur. Phys. J.} \textbf{\bibinfo{volume}{C25}},
  \bibinfo{pages}{77} (\bibinfo{year}{2002}), \eprint{hep-ph/0204115}.

\bibitem[{\citenamefont{Andersen et~al.}(2004)}]{Andersen:2003xj}
\bibinfo{author}{\bibfnamefont{J.~R.} \bibnamefont{Andersen}}
  \bibnamefont{et~al.} (\bibinfo{collaboration}{Small x}),
  \bibinfo{journal}{Eur. Phys. J.} \textbf{\bibinfo{volume}{C35}},
  \bibinfo{pages}{67} (\bibinfo{year}{2004}), \eprint{hep-ph/0312333}.

\bibitem[{\citenamefont{Andersen et~al.}(2006)}]{Andersen:2006pg}
\bibinfo{author}{\bibfnamefont{J.~R.} \bibnamefont{Andersen}}
  \bibnamefont{et~al.} (\bibinfo{collaboration}{Small x}),
  \bibinfo{journal}{Eur. Phys. J.} \textbf{\bibinfo{volume}{C48}},
  \bibinfo{pages}{53} (\bibinfo{year}{2006}), \eprint{hep-ph/0604189}.

\bibitem[{\citenamefont{Luszczak and
  Szczurek}(2006{\natexlab{b}})}]{Luszczak:2005xs}
\bibinfo{author}{\bibfnamefont{M.}~\bibnamefont{Luszczak}} \bibnamefont{and}
  \bibinfo{author}{\bibfnamefont{A.}~\bibnamefont{Szczurek}},
  \bibinfo{journal}{Eur. Phys. J.} \textbf{\bibinfo{volume}{C46}},
  \bibinfo{pages}{123} (\bibinfo{year}{2006}{\natexlab{b}}),
  \eprint{hep-ph/0504119}.

\bibitem[{\citenamefont{Lipatov and Zotov}(2005)}]{Lipatov:2005at}
\bibinfo{author}{\bibfnamefont{A.~V.} \bibnamefont{Lipatov}} \bibnamefont{and}
  \bibinfo{author}{\bibfnamefont{N.~P.} \bibnamefont{Zotov}},
  \bibinfo{journal}{Eur. Phys. J.} \textbf{\bibinfo{volume}{C44}},
  \bibinfo{pages}{559} (\bibinfo{year}{2005}), \eprint{hep-ph/0501172}.

\bibitem[{\citenamefont{Iancu et~al.}(2002)\citenamefont{Iancu, Leonidov, and
  McLerran}}]{Iancu:2002xk}
\bibinfo{author}{\bibfnamefont{E.}~\bibnamefont{Iancu}},
  \bibinfo{author}{\bibfnamefont{A.}~\bibnamefont{Leonidov}}, \bibnamefont{and}
  \bibinfo{author}{\bibfnamefont{L.}~\bibnamefont{McLerran}}
  (\bibinfo{year}{2002}), \eprint{hep-ph/0202270}.

\bibitem[{\citenamefont{Iancu and Venugopalan}(2003)}]{Iancu:2003xm}
\bibinfo{author}{\bibfnamefont{E.}~\bibnamefont{Iancu}} \bibnamefont{and}
  \bibinfo{author}{\bibfnamefont{R.}~\bibnamefont{Venugopalan}}
  (\bibinfo{year}{2003}), \eprint{hep-ph/0303204}.

\bibitem[{\citenamefont{Yao et~al.}(2006)}]{Yao:2006px}
\bibinfo{author}{\bibfnamefont{W.~M.} \bibnamefont{Yao}} \bibnamefont{et~al.}
  (\bibinfo{collaboration}{Particle Data Group}), \bibinfo{journal}{J. Phys.}
  \textbf{\bibinfo{volume}{G33}}, \bibinfo{pages}{1} (\bibinfo{year}{2006}).

\bibitem[{\citenamefont{'t~Hooft}(1976)}]{Hooft:1976up}
\bibinfo{author}{\bibfnamefont{G.}~\bibnamefont{'t~Hooft}},
  \bibinfo{journal}{Phys. Rev. Lett.} \textbf{\bibinfo{volume}{37}},
  \bibinfo{pages}{8} (\bibinfo{year}{1976}).

\bibitem[{\citenamefont{Witten}(1979)}]{Witten:1979vv}
\bibinfo{author}{\bibfnamefont{E.}~\bibnamefont{Witten}},
  \bibinfo{journal}{Nucl. Phys.} \textbf{\bibinfo{volume}{B156}},
  \bibinfo{pages}{269} (\bibinfo{year}{1979}).

\bibitem[{\citenamefont{Atwood and Soni}(1997)}]{Atwood:1997bn}
\bibinfo{author}{\bibfnamefont{D.}~\bibnamefont{Atwood}} \bibnamefont{and}
  \bibinfo{author}{\bibfnamefont{A.}~\bibnamefont{Soni}},
  \bibinfo{journal}{Phys. Lett.} \textbf{\bibinfo{volume}{B405}},
  \bibinfo{pages}{150} (\bibinfo{year}{1997}), \eprint{hep-ph/9704357}.

\bibitem[{\citenamefont{Ali and Parkhomenko}(2003)}]{Ali:2003kg}
\bibinfo{author}{\bibfnamefont{A.}~\bibnamefont{Ali}} \bibnamefont{and}
  \bibinfo{author}{\bibfnamefont{A.~Y.} \bibnamefont{Parkhomenko}},
  \bibinfo{journal}{Eur. Phys. J.} \textbf{\bibinfo{volume}{C30}},
  \bibinfo{pages}{367} (\bibinfo{year}{2003}), \eprint{hep-ph/0307092}.

\bibitem[{\citenamefont{Ali and Parkhomenko}(2002)}]{Ali:2000ci}
\bibinfo{author}{\bibfnamefont{A.}~\bibnamefont{Ali}} \bibnamefont{and}
  \bibinfo{author}{\bibfnamefont{A.~Y.} \bibnamefont{Parkhomenko}},
  \bibinfo{journal}{Phys. Rev.} \textbf{\bibinfo{volume}{D65}},
  \bibinfo{pages}{074020} (\bibinfo{year}{2002}), \eprint{hep-ph/0012212}.

\bibitem[{\citenamefont{Muta and Yang}(2000)}]{Muta:1999tc}
\bibinfo{author}{\bibfnamefont{T.}~\bibnamefont{Muta}} \bibnamefont{and}
  \bibinfo{author}{\bibfnamefont{M.-Z.} \bibnamefont{Yang}},
  \bibinfo{journal}{Phys. Rev.} \textbf{\bibinfo{volume}{D61}},
  \bibinfo{pages}{054007} (\bibinfo{year}{2000}), \eprint{hep-ph/9909484}.

\bibitem[{\citenamefont{Ahmady et~al.}(1998{\natexlab{a}})\citenamefont{Ahmady,
  Elias, and Kou}}]{Ahmady:1998mi}
\bibinfo{author}{\bibfnamefont{M.~R.} \bibnamefont{Ahmady}},
  \bibinfo{author}{\bibfnamefont{V.}~\bibnamefont{Elias}}, \bibnamefont{and}
  \bibinfo{author}{\bibfnamefont{E.}~\bibnamefont{Kou}},
  \bibinfo{journal}{Phys. Rev.} \textbf{\bibinfo{volume}{D57}},
  \bibinfo{pages}{7034} (\bibinfo{year}{1998}{\natexlab{a}}),
  \eprint{hep-ph/9801447}.

\bibitem[{\citenamefont{Agaev and Stefanis}(2004)}]{Agaev:2002ek}
\bibinfo{author}{\bibfnamefont{S.~S.} \bibnamefont{Agaev}} \bibnamefont{and}
  \bibinfo{author}{\bibfnamefont{N.~G.} \bibnamefont{Stefanis}},
  \bibinfo{journal}{Eur. Phys. J.} \textbf{\bibinfo{volume}{C32}},
  \bibinfo{pages}{507} (\bibinfo{year}{2004}), \eprint{hep-ph/0212318}.

\bibitem[{\citenamefont{Kroll and Passek-Kumericki}(2003)}]{Kroll:2002nt}
\bibinfo{author}{\bibfnamefont{P.}~\bibnamefont{Kroll}} \bibnamefont{and}
  \bibinfo{author}{\bibfnamefont{K.}~\bibnamefont{Passek-Kumericki}},
  \bibinfo{journal}{Phys. Rev.} \textbf{\bibinfo{volume}{D67}},
  \bibinfo{pages}{054017} (\bibinfo{year}{2003}), \eprint{hep-ph/0210045}.

\bibitem[{\citenamefont{Szczurek et~al.}(2007)\citenamefont{Szczurek,
  Pasechnik, and Teryaev}}]{Szczurek:2006bn}
\bibinfo{author}{\bibfnamefont{A.}~\bibnamefont{Szczurek}},
  \bibinfo{author}{\bibfnamefont{R.~S.} \bibnamefont{Pasechnik}},
  \bibnamefont{and} \bibinfo{author}{\bibfnamefont{O.~V.}
  \bibnamefont{Teryaev}}, \bibinfo{journal}{Phys. Rev.}
  \textbf{\bibinfo{volume}{D75}}, \bibinfo{pages}{054021}
  (\bibinfo{year}{2007}), \eprint{hep-ph/0608302}.

\bibitem[{\citenamefont{Jalilian-Marian and
  Jeon}(2002)}]{JalilianMarian:2001bu}
\bibinfo{author}{\bibfnamefont{J.}~\bibnamefont{Jalilian-Marian}}
  \bibnamefont{and} \bibinfo{author}{\bibfnamefont{S.}~\bibnamefont{Jeon}},
  \bibinfo{journal}{Phys. Rev.} \textbf{\bibinfo{volume}{C65}},
  \bibinfo{pages}{065201} (\bibinfo{year}{2002}), \eprint{hep-ph/0110417}.

\bibitem[{\citenamefont{Dumitru
  et~al.}(2006{\natexlab{a}})\citenamefont{Dumitru, Hayashigaki, and
  Jalilian-Marian}}]{Dumitru:2005gt}
\bibinfo{author}{\bibfnamefont{A.}~\bibnamefont{Dumitru}},
  \bibinfo{author}{\bibfnamefont{A.}~\bibnamefont{Hayashigaki}},
  \bibnamefont{and}
  \bibinfo{author}{\bibfnamefont{J.}~\bibnamefont{Jalilian-Marian}},
  \bibinfo{journal}{Nucl. Phys.} \textbf{\bibinfo{volume}{A765}},
  \bibinfo{pages}{464} (\bibinfo{year}{2006}{\natexlab{a}}),
  \eprint{hep-ph/0506308}.

\bibitem[{\citenamefont{Dumitru
  et~al.}(2006{\natexlab{b}})\citenamefont{Dumitru, Hayashigaki, and
  Jalilian-Marian}}]{Dumitru:2005kb}
\bibinfo{author}{\bibfnamefont{A.}~\bibnamefont{Dumitru}},
  \bibinfo{author}{\bibfnamefont{A.}~\bibnamefont{Hayashigaki}},
  \bibnamefont{and}
  \bibinfo{author}{\bibfnamefont{J.}~\bibnamefont{Jalilian-Marian}},
  \bibinfo{journal}{Nucl. Phys.} \textbf{\bibinfo{volume}{A770}},
  \bibinfo{pages}{57} (\bibinfo{year}{2006}{\natexlab{b}}),
  \eprint{hep-ph/0512129}.

\bibitem[{\citenamefont{Tuchin}(2008)}]{Tuchin:2007pf}
\bibinfo{author}{\bibfnamefont{K.}~\bibnamefont{Tuchin}},
  \bibinfo{journal}{Nucl. Phys.} \textbf{\bibinfo{volume}{A798}},
  \bibinfo{pages}{61} (\bibinfo{year}{2008}), \eprint{0705.2193}.

\bibitem[{\citenamefont{Fillion-Gourdeau and
  Jeon}(2008)}]{FillionGourdeau:2007ee}
\bibinfo{author}{\bibfnamefont{F.}~\bibnamefont{Fillion-Gourdeau}}
  \bibnamefont{and} \bibinfo{author}{\bibfnamefont{S.}~\bibnamefont{Jeon}},
  \bibinfo{journal}{Phys. Rev.} \textbf{\bibinfo{volume}{C77}},
  \bibinfo{pages}{055201} (\bibinfo{year}{2008}), \eprint{0709.4196}.

\bibitem[{\citenamefont{Kagan and Petrov}(1997)}]{Kagan:1997qn}
\bibinfo{author}{\bibfnamefont{A.~L.} \bibnamefont{Kagan}} \bibnamefont{and}
  \bibinfo{author}{\bibfnamefont{A.~A.} \bibnamefont{Petrov}}
  (\bibinfo{year}{1997}), \eprint{hep-ph/9707354}.

\bibitem[{\citenamefont{Hou and Tseng}(1998)}]{Hou:1997wy}
\bibinfo{author}{\bibfnamefont{W.-S.} \bibnamefont{Hou}} \bibnamefont{and}
  \bibinfo{author}{\bibfnamefont{B.}~\bibnamefont{Tseng}},
  \bibinfo{journal}{Phys. Rev. Lett.} \textbf{\bibinfo{volume}{80}},
  \bibinfo{pages}{434} (\bibinfo{year}{1998}), \eprint{hep-ph/9705304}.

\bibitem[{\citenamefont{Ahmady et~al.}(1998{\natexlab{b}})\citenamefont{Ahmady,
  Kou, and Sugamoto}}]{Ahmady:1997fa}
\bibinfo{author}{\bibfnamefont{M.~R.} \bibnamefont{Ahmady}},
  \bibinfo{author}{\bibfnamefont{E.}~\bibnamefont{Kou}}, \bibnamefont{and}
  \bibinfo{author}{\bibfnamefont{A.}~\bibnamefont{Sugamoto}},
  \bibinfo{journal}{Phys. Rev.} \textbf{\bibinfo{volume}{D58}},
  \bibinfo{pages}{014015} (\bibinfo{year}{1998}{\natexlab{b}}),
  \eprint{hep-ph/9710509}.

\bibitem[{\citenamefont{Du et~al.}(1998)\citenamefont{Du, Kim, and
  Yang}}]{Du:1997hs}
\bibinfo{author}{\bibfnamefont{D.-s.} \bibnamefont{Du}},
  \bibinfo{author}{\bibfnamefont{C.~S.} \bibnamefont{Kim}}, \bibnamefont{and}
  \bibinfo{author}{\bibfnamefont{Y.-d.} \bibnamefont{Yang}},
  \bibinfo{journal}{Phys. Lett.} \textbf{\bibinfo{volume}{B426}},
  \bibinfo{pages}{133} (\bibinfo{year}{1998}), \eprint{hep-ph/9711428}.

\bibitem[{\citenamefont{Jeon and Jalilian-Marian}(2002)}]{Jeon:2002hh}
\bibinfo{author}{\bibfnamefont{S.}~\bibnamefont{Jeon}} \bibnamefont{and}
  \bibinfo{author}{\bibfnamefont{J.}~\bibnamefont{Jalilian-Marian}},
  \bibinfo{journal}{Nucl. Phys.} \textbf{\bibinfo{volume}{A710}},
  \bibinfo{pages}{145} (\bibinfo{year}{2002}), \eprint{hep-ph/0203105}.

\bibitem[{\citenamefont{Jeon}(2002)}]{Jeon:2001cy}
\bibinfo{author}{\bibfnamefont{S.}~\bibnamefont{Jeon}}, \bibinfo{journal}{Phys.
  Rev.} \textbf{\bibinfo{volume}{C65}}, \bibinfo{pages}{024903}
  (\bibinfo{year}{2002}), \eprint{hep-ph/0107140}.

\bibitem[{\citenamefont{Venugopalan}(2004)}]{Venugopalan:2004dj}
\bibinfo{author}{\bibfnamefont{R.}~\bibnamefont{Venugopalan}}
  (\bibinfo{year}{2004}), \eprint{hep-ph/0412396}.

\bibitem[{\citenamefont{McLerran and
  Venugopalan}(1994{\natexlab{a}})}]{McLerran:1993ka}
\bibinfo{author}{\bibfnamefont{L.~D.} \bibnamefont{McLerran}} \bibnamefont{and}
  \bibinfo{author}{\bibfnamefont{R.}~\bibnamefont{Venugopalan}},
  \bibinfo{journal}{Phys. Rev.} \textbf{\bibinfo{volume}{D49}},
  \bibinfo{pages}{3352} (\bibinfo{year}{1994}{\natexlab{a}}),
  \eprint{hep-ph/9311205}.

\bibitem[{\citenamefont{McLerran and
  Venugopalan}(1994{\natexlab{b}})}]{McLerran:1993ni}
\bibinfo{author}{\bibfnamefont{L.~D.} \bibnamefont{McLerran}} \bibnamefont{and}
  \bibinfo{author}{\bibfnamefont{R.}~\bibnamefont{Venugopalan}},
  \bibinfo{journal}{Phys. Rev.} \textbf{\bibinfo{volume}{D49}},
  \bibinfo{pages}{2233} (\bibinfo{year}{1994}{\natexlab{b}}),
  \eprint{hep-ph/9309289}.

\bibitem[{\citenamefont{Iancu et~al.}(2001)\citenamefont{Iancu, Leonidov, and
  McLerran}}]{Iancu:2000hn}
\bibinfo{author}{\bibfnamefont{E.}~\bibnamefont{Iancu}},
  \bibinfo{author}{\bibfnamefont{A.}~\bibnamefont{Leonidov}}, \bibnamefont{and}
  \bibinfo{author}{\bibfnamefont{L.~D.} \bibnamefont{McLerran}},
  \bibinfo{journal}{Nucl. Phys.} \textbf{\bibinfo{volume}{A692}},
  \bibinfo{pages}{583} (\bibinfo{year}{2001}), \eprint{hep-ph/0011241}.

\bibitem[{\citenamefont{Ferreiro et~al.}(2002)\citenamefont{Ferreiro, Iancu,
  Leonidov, and McLerran}}]{Ferreiro:2001qy}
\bibinfo{author}{\bibfnamefont{E.}~\bibnamefont{Ferreiro}},
  \bibinfo{author}{\bibfnamefont{E.}~\bibnamefont{Iancu}},
  \bibinfo{author}{\bibfnamefont{A.}~\bibnamefont{Leonidov}}, \bibnamefont{and}
  \bibinfo{author}{\bibfnamefont{L.}~\bibnamefont{McLerran}},
  \bibinfo{journal}{Nucl. Phys.} \textbf{\bibinfo{volume}{A703}},
  \bibinfo{pages}{489} (\bibinfo{year}{2002}), \eprint{hep-ph/0109115}.

\bibitem[{\citenamefont{Jalilian-Marian
  et~al.}(1997{\natexlab{a}})\citenamefont{Jalilian-Marian, Kovner, McLerran,
  and Weigert}}]{JalilianMarian:1996xn}
\bibinfo{author}{\bibfnamefont{J.}~\bibnamefont{Jalilian-Marian}},
  \bibinfo{author}{\bibfnamefont{A.}~\bibnamefont{Kovner}},
  \bibinfo{author}{\bibfnamefont{L.~D.} \bibnamefont{McLerran}},
  \bibnamefont{and} \bibinfo{author}{\bibfnamefont{H.}~\bibnamefont{Weigert}},
  \bibinfo{journal}{Phys. Rev.} \textbf{\bibinfo{volume}{D55}},
  \bibinfo{pages}{5414} (\bibinfo{year}{1997}{\natexlab{a}}),
  \eprint{hep-ph/9606337}.

\bibitem[{\citenamefont{Jalilian-Marian
  et~al.}(1999)\citenamefont{Jalilian-Marian, Kovner, Leonidov, and
  Weigert}}]{JalilianMarian:1997gr}
\bibinfo{author}{\bibfnamefont{J.}~\bibnamefont{Jalilian-Marian}},
  \bibinfo{author}{\bibfnamefont{A.}~\bibnamefont{Kovner}},
  \bibinfo{author}{\bibfnamefont{A.}~\bibnamefont{Leonidov}}, \bibnamefont{and}
  \bibinfo{author}{\bibfnamefont{H.}~\bibnamefont{Weigert}},
  \bibinfo{journal}{Phys. Rev.} \textbf{\bibinfo{volume}{D59}},
  \bibinfo{pages}{014014} (\bibinfo{year}{1999}), \eprint{hep-ph/9706377}.

\bibitem[{\citenamefont{Jalilian-Marian
  et~al.}(1997{\natexlab{b}})\citenamefont{Jalilian-Marian, Kovner, Leonidov,
  and Weigert}}]{JalilianMarian:1997jx}
\bibinfo{author}{\bibfnamefont{J.}~\bibnamefont{Jalilian-Marian}},
  \bibinfo{author}{\bibfnamefont{A.}~\bibnamefont{Kovner}},
  \bibinfo{author}{\bibfnamefont{A.}~\bibnamefont{Leonidov}}, \bibnamefont{and}
  \bibinfo{author}{\bibfnamefont{H.}~\bibnamefont{Weigert}},
  \bibinfo{journal}{Nucl. Phys.} \textbf{\bibinfo{volume}{B504}},
  \bibinfo{pages}{415} (\bibinfo{year}{1997}{\natexlab{b}}),
  \eprint{hep-ph/9701284}.

\bibitem[{\citenamefont{Baltz et~al.}(2001)\citenamefont{Baltz, Gelis,
  McLerran, and Peshier}}]{Baltz:2001dp}
\bibinfo{author}{\bibfnamefont{A.~J.} \bibnamefont{Baltz}},
  \bibinfo{author}{\bibfnamefont{F.}~\bibnamefont{Gelis}},
  \bibinfo{author}{\bibfnamefont{L.~D.} \bibnamefont{McLerran}},
  \bibnamefont{and} \bibinfo{author}{\bibfnamefont{A.}~\bibnamefont{Peshier}},
  \bibinfo{journal}{Nucl. Phys.} \textbf{\bibinfo{volume}{A695}},
  \bibinfo{pages}{395} (\bibinfo{year}{2001}), \eprint{nucl-th/0101024}.

\bibitem[{\citenamefont{Itzykson and Zuber}(1980)}]{Itzykson:1980rh}
\bibinfo{author}{\bibfnamefont{C.}~\bibnamefont{Itzykson}} \bibnamefont{and}
  \bibinfo{author}{\bibfnamefont{J.~B.} \bibnamefont{Zuber}}
  (\bibinfo{year}{1980}), \bibinfo{note}{new York, Usa: Mcgraw-hill (1980) 705
  P.(International Series In Pure and Applied Physics)}.

\bibitem[{\citenamefont{Gelis and
  Venugopalan}(2004{\natexlab{a}})}]{Gelis:2003vh}
\bibinfo{author}{\bibfnamefont{F.}~\bibnamefont{Gelis}} \bibnamefont{and}
  \bibinfo{author}{\bibfnamefont{R.}~\bibnamefont{Venugopalan}},
  \bibinfo{journal}{Phys. Rev.} \textbf{\bibinfo{volume}{D69}},
  \bibinfo{pages}{014019} (\bibinfo{year}{2004}{\natexlab{a}}),
  \eprint{hep-ph/0310090}.

\bibitem[{\citenamefont{Gelis and
  Venugopalan}(2004{\natexlab{b}})}]{Gelis:2004bz}
\bibinfo{author}{\bibfnamefont{F.}~\bibnamefont{Gelis}} \bibnamefont{and}
  \bibinfo{author}{\bibfnamefont{R.}~\bibnamefont{Venugopalan}},
  \bibinfo{journal}{J. Phys.} \textbf{\bibinfo{volume}{G30}},
  \bibinfo{pages}{S995} (\bibinfo{year}{2004}{\natexlab{b}}),
  \eprint{hep-ph/0403229}.

\bibitem[{\citenamefont{Blaizot
  et~al.}(2004{\natexlab{a}})\citenamefont{Blaizot, Gelis, and
  Venugopalan}}]{Blaizot:2004wu}
\bibinfo{author}{\bibfnamefont{J.~P.} \bibnamefont{Blaizot}},
  \bibinfo{author}{\bibfnamefont{F.}~\bibnamefont{Gelis}}, \bibnamefont{and}
  \bibinfo{author}{\bibfnamefont{R.}~\bibnamefont{Venugopalan}},
  \bibinfo{journal}{Nucl. Phys.} \textbf{\bibinfo{volume}{A743}},
  \bibinfo{pages}{13} (\bibinfo{year}{2004}{\natexlab{a}}),
  \eprint{hep-ph/0402256}.

\bibitem[{\citenamefont{Dumitru and McLerran}(2002)}]{Dumitru:2001ux}
\bibinfo{author}{\bibfnamefont{A.}~\bibnamefont{Dumitru}} \bibnamefont{and}
  \bibinfo{author}{\bibfnamefont{L.~D.} \bibnamefont{McLerran}},
  \bibinfo{journal}{Nucl. Phys.} \textbf{\bibinfo{volume}{A700}},
  \bibinfo{pages}{492} (\bibinfo{year}{2002}), \eprint{hep-ph/0105268}.

\bibitem[{\citenamefont{Gelis and Mehtar-Tani}(2006)}]{Gelis:2005pt}
\bibinfo{author}{\bibfnamefont{F.}~\bibnamefont{Gelis}} \bibnamefont{and}
  \bibinfo{author}{\bibfnamefont{Y.}~\bibnamefont{Mehtar-Tani}},
  \bibinfo{journal}{Phys. Rev.} \textbf{\bibinfo{volume}{D73}},
  \bibinfo{pages}{034019} (\bibinfo{year}{2006}), \eprint{hep-ph/0512079}.

\bibitem[{\citenamefont{Kovchegov and Mueller}(1998)}]{Kovchegov:1998bi}
\bibinfo{author}{\bibfnamefont{Y.~V.} \bibnamefont{Kovchegov}}
  \bibnamefont{and} \bibinfo{author}{\bibfnamefont{A.~H.}
  \bibnamefont{Mueller}}, \bibinfo{journal}{Nucl. Phys.}
  \textbf{\bibinfo{volume}{B529}}, \bibinfo{pages}{451} (\bibinfo{year}{1998}),
  \eprint{hep-ph/9802440}.

\bibitem[{\citenamefont{Fukushima and Hidaka}(2008)}]{Fukushima:2008ya}
\bibinfo{author}{\bibfnamefont{K.}~\bibnamefont{Fukushima}} \bibnamefont{and}
  \bibinfo{author}{\bibfnamefont{Y.}~\bibnamefont{Hidaka}}
  (\bibinfo{year}{2008}), \eprint{0806.2143}.

\bibitem[{\citenamefont{Gelis and Venugopalan}(2006)}]{Gelis:2006yv}
\bibinfo{author}{\bibfnamefont{F.}~\bibnamefont{Gelis}} \bibnamefont{and}
  \bibinfo{author}{\bibfnamefont{R.}~\bibnamefont{Venugopalan}},
  \bibinfo{journal}{Nucl. Phys.} \textbf{\bibinfo{volume}{A776}},
  \bibinfo{pages}{135} (\bibinfo{year}{2006}), \eprint{hep-ph/0601209}.

\bibitem[{\citenamefont{Gyulassy and McLerran}(1997)}]{Gyulassy:1997vt}
\bibinfo{author}{\bibfnamefont{M.}~\bibnamefont{Gyulassy}} \bibnamefont{and}
  \bibinfo{author}{\bibfnamefont{L.~D.} \bibnamefont{McLerran}},
  \bibinfo{journal}{Phys. Rev.} \textbf{\bibinfo{volume}{C56}},
  \bibinfo{pages}{2219} (\bibinfo{year}{1997}), \eprint{nucl-th/9704034}.

\bibitem[{\citenamefont{Blaizot
  et~al.}(2004{\natexlab{b}})\citenamefont{Blaizot, Gelis, and
  Venugopalan}}]{Blaizot:2004wv}
\bibinfo{author}{\bibfnamefont{J.~P.} \bibnamefont{Blaizot}},
  \bibinfo{author}{\bibfnamefont{F.}~\bibnamefont{Gelis}}, \bibnamefont{and}
  \bibinfo{author}{\bibfnamefont{R.}~\bibnamefont{Venugopalan}},
  \bibinfo{journal}{Nucl. Phys.} \textbf{\bibinfo{volume}{A743}},
  \bibinfo{pages}{57} (\bibinfo{year}{2004}{\natexlab{b}}),
  \eprint{hep-ph/0402257}.

\bibitem[{\citenamefont{Gelis and Peshier}(2002)}]{Gelis:2001da}
\bibinfo{author}{\bibfnamefont{F.}~\bibnamefont{Gelis}} \bibnamefont{and}
  \bibinfo{author}{\bibfnamefont{A.}~\bibnamefont{Peshier}},
  \bibinfo{journal}{Nucl. Phys.} \textbf{\bibinfo{volume}{A697}},
  \bibinfo{pages}{879} (\bibinfo{year}{2002}), \eprint{hep-ph/0107142}.

\bibitem[{\citenamefont{Fukushima and Hidaka}(2007)}]{Fukushima:2007dy}
\bibinfo{author}{\bibfnamefont{K.}~\bibnamefont{Fukushima}} \bibnamefont{and}
  \bibinfo{author}{\bibfnamefont{Y.}~\bibnamefont{Hidaka}},
  \bibinfo{journal}{JHEP} \textbf{\bibinfo{volume}{06}}, \bibinfo{pages}{040}
  (\bibinfo{year}{2007}), \eprint{0704.2806}.

\bibitem[{\citenamefont{Ciafaloni}(1988)}]{Ciafaloni:1987ur}
\bibinfo{author}{\bibfnamefont{M.}~\bibnamefont{Ciafaloni}},
  \bibinfo{journal}{Nucl. Phys.} \textbf{\bibinfo{volume}{B296}},
  \bibinfo{pages}{49} (\bibinfo{year}{1988}).

\bibitem[{\citenamefont{Catani et~al.}(1990{\natexlab{a}})\citenamefont{Catani,
  Fiorani, and Marchesini}}]{Catani:1989sg}
\bibinfo{author}{\bibfnamefont{S.}~\bibnamefont{Catani}},
  \bibinfo{author}{\bibfnamefont{F.}~\bibnamefont{Fiorani}}, \bibnamefont{and}
  \bibinfo{author}{\bibfnamefont{G.}~\bibnamefont{Marchesini}},
  \bibinfo{journal}{Nucl. Phys.} \textbf{\bibinfo{volume}{B336}},
  \bibinfo{pages}{18} (\bibinfo{year}{1990}{\natexlab{a}}).

\bibitem[{\citenamefont{Catani et~al.}(1990{\natexlab{b}})\citenamefont{Catani,
  Fiorani, and Marchesini}}]{Catani:1989yc}
\bibinfo{author}{\bibfnamefont{S.}~\bibnamefont{Catani}},
  \bibinfo{author}{\bibfnamefont{F.}~\bibnamefont{Fiorani}}, \bibnamefont{and}
  \bibinfo{author}{\bibfnamefont{G.}~\bibnamefont{Marchesini}},
  \bibinfo{journal}{Phys. Lett.} \textbf{\bibinfo{volume}{B234}},
  \bibinfo{pages}{339} (\bibinfo{year}{1990}{\natexlab{b}}).

\bibitem[{\citenamefont{Jung and Salam}(2001)}]{Jung:2000hk}
\bibinfo{author}{\bibfnamefont{H.}~\bibnamefont{Jung}} \bibnamefont{and}
  \bibinfo{author}{\bibfnamefont{G.~P.} \bibnamefont{Salam}},
  \bibinfo{journal}{Eur. Phys. J.} \textbf{\bibinfo{volume}{C19}},
  \bibinfo{pages}{351} (\bibinfo{year}{2001}), \eprint{hep-ph/0012143}.

\bibitem[{\citenamefont{Jung}(2004)}]{Jung:2003wu}
\bibinfo{author}{\bibfnamefont{H.}~\bibnamefont{Jung}}, \bibinfo{journal}{Mod.
  Phys. Lett.} \textbf{\bibinfo{volume}{A19}}, \bibinfo{pages}{1}
  (\bibinfo{year}{2004}), \eprint{hep-ph/0311249}.

\bibitem[{\citenamefont{Kimber et~al.}(2001)\citenamefont{Kimber, Martin, and
  Ryskin}}]{Kimber:2001sc}
\bibinfo{author}{\bibfnamefont{M.~A.} \bibnamefont{Kimber}},
  \bibinfo{author}{\bibfnamefont{A.~D.} \bibnamefont{Martin}},
  \bibnamefont{and} \bibinfo{author}{\bibfnamefont{M.~G.}
  \bibnamefont{Ryskin}}, \bibinfo{journal}{Phys. Rev.}
  \textbf{\bibinfo{volume}{D63}}, \bibinfo{pages}{114027}
  (\bibinfo{year}{2001}), \eprint{hep-ph/0101348}.

\bibitem[{\citenamefont{Hahn}(2005)}]{Hahn:2004fe}
\bibinfo{author}{\bibfnamefont{T.}~\bibnamefont{Hahn}},
  \bibinfo{journal}{Comput. Phys. Commun.} \textbf{\bibinfo{volume}{168}},
  \bibinfo{pages}{78} (\bibinfo{year}{2005}), \eprint{hep-ph/0404043}.

\bibitem[{\citenamefont{Lam and Mahlon}(2000)}]{Lam:1999wu}
\bibinfo{author}{\bibfnamefont{C.~S.} \bibnamefont{Lam}} \bibnamefont{and}
  \bibinfo{author}{\bibfnamefont{G.}~\bibnamefont{Mahlon}},
  \bibinfo{journal}{Phys. Rev.} \textbf{\bibinfo{volume}{D61}},
  \bibinfo{pages}{014005} (\bibinfo{year}{2000}), \eprint{hep-ph/9907281}.

\bibitem[{\citenamefont{Kovchegov and Rischke}(1997)}]{Kovchegov:1997ke}
\bibinfo{author}{\bibfnamefont{Y.~V.} \bibnamefont{Kovchegov}}
  \bibnamefont{and} \bibinfo{author}{\bibfnamefont{D.~H.}
  \bibnamefont{Rischke}}, \bibinfo{journal}{Phys. Rev.}
  \textbf{\bibinfo{volume}{C56}}, \bibinfo{pages}{1084} (\bibinfo{year}{1997}),
  \eprint{hep-ph/9704201}.

\end{thebibliography}

\end{document}